\documentclass[10pt, conference, compsocconf]{IEEEtran}
\IEEEoverridecommandlockouts
\usepackage{graphicx}
\usepackage{amssymb}
\usepackage{balance}
\usepackage{amsmath}
\usepackage{subfig}
\usepackage{url}
\usepackage[usenames,dvipsnames]{pstricks}
\usepackage{pstricks-add}
\usepackage{pstricks,pst-node,pst-tree}
\usepackage{multirow} 
\usepackage{wrapfig}
\newcommand{\alg}{{\sc MiRage}}

\def\qed{\rule{2mm}{2mm}}

  \renewenvironment{thebibliography}[1]{%
    \begin{oldthebibliography}{#1}%
      \setlength{\parskip}{0ex}%
      \setlength{\itemsep}{0ex}%
  }%
  {%
    \end{oldthebibliography}%
  }

\begin{document}


\title{\alg: An Iterative MapReduce based Frequent Subgraph Mining Algorithm}

\author{
Mansurul A Bhuiyan, and Mohammad Al Hasan \\
{Dept. of Computer Science, Indiana University---Purdue University, Indianapolis}\\
{\{mbhuiyan, alhasan\}@cs.iupui.edu}
}
\maketitle

\begin{abstract}

Frequent subgraph mining (FSM) is an important task for exploratory data
analysis on graph data.  Over the years, many algorithms have been proposed to
solve this task. These algorithms assume that the data structure of the mining
task is small enough to fit in the main memory of a computer.  However, as the
real-world graph data grows, both in size and quantity, such an assumption does
not hold any longer. To overcome this, some graph database-centric methods have
been proposed in recent years for solving FSM; however, a distributed solution
using MapReduce paradigm has not been explored extensively. Since, MapReduce
is becoming the de-facto paradigm for computation on massive data,
an efficient FSM algorithm on this paradigm is of huge demand.  In this work, we
propose a frequent subgraph mining algorithm called \alg\ which uses an
iterative MapReduce based framework.  \alg\ is complete as it returns all the
frequent subgraphs for a given user-defined support, and it is efficient as it
applies all the optimizations that the latest FSM algorithms adopt.  Our
experiments with real life and large synthetic datasets validate the effectiveness 
of \alg\ for mining frequent subgraphs from large graph datasets. 
The source code of \alg\ is available from \url{www.cs.iupui.edu/~alhasan/software/}

\end{abstract}

\section{Introduction}

In recent years, the ``big data'' phenomenon has engulfed a significant number
of research and application domains including data mining,
computational biology~\cite{Rosenthal.Mork.ea:10}, environmental sciences,
e-commerce~\cite{Lam.Liu.ea:2012}, web mining, and social network
analysis~\cite{Liu.Zhang.ea:2010}. In these domains, analyzing and mining of
massive data for extracting novel insights has become a routine task.  However,
traditional methods for data analysis and mining are not designed to handle
massive amount of data, so in recent years many such methods are re-designed
and re-implemented under a computing framework that is aware of the big-data
syndrome.

Among the recent efforts for building a suitable computing platform for
analyzing massive data, the MapReduce~\cite{Dean.Ghemawat:2008} framework of
distributed computing has been the most successful. It adopts a data centric
approach of distributed computing with the ideology of ``moving computation to
data''; besides it uses a distributed file system that is particularly
optimized to improve the IO performance while handling massive
data. Another main reason for this framework to gain attention of many admirers
is the higher level of abstraction that it provides, which keeps many system
level details hidden from the programmers and allows them to concentrate more on
the problem specific computational logic.

MapReduce has become a popular platform for analyzing large networks in recent
years. However, the majority of such analyses are limited to estimating global
statistics (such as diameter)~\cite{Kang.Tsourakakis.ea:2009}, spectral
analysis~\cite{Kang.Meeder:ea:11}, or vertex-centrality
analysis~\cite{Kang.Tsourakakis.ea:2009}. Efforts for mining sub-structure is
not that common, except a few works for counting
triangles~\cite{Suri.Sergei.ea:2011}. Specifically,
frequent subgraph mining on MapReduce has received the least attention. Given
the growth of applications of frequent subgraph mining in various disciplines
including social networks, bioinformatics~\cite{Huan.Wang.ea:2004},
cheminformatics~\cite{Kramer.Raedt.ea:2004}, and semantic
web~\cite{Berendt:2006}, a scalable method for frequent subgraph mining on
MapReduce is of high demand.

Solving the task of frequent subgraph mining (FSM) on a distributed platform
like MapReduce is challenging for various reasons. First, an FSM method
computes the support of a candidate subgraph pattern over the entire set of
input graphs in a graph dataset. In a distributed platform, if the input
graphs are partitioned over various worker nodes, the local support of a
subgraph at a worker node is not much useful for deciding whether the given
subgraph is frequent or not. Also, local support of a subgraph in various nodes cannot be
aggregated in a global data structure, because,  MapReduce programming model
does not provide any built-in mechanism for communicating with a global state.
Also, the support computation cannot be delayed arbitrarily, as following Apriori
principle~\cite{Agrawal.Srikant:1994}, future
candidate frequent patterns~\footnote{In data mining, the word {\em pattern} is
generally used to denote a combinatorial object, such as a set, a sequence, a
tree or a graph. In this paper, we use pattern to denote a graph object only.}
can be generated only from a frequent pattern. 

In this paper, we propose, \alg~\footnote{\alg~is an anagram of the bold letters
in {\bf I}terative {\bf M}ap {\bf R}educe based sub{\bf G}gr{\bf A}ph {\bf
E}xtraction}, a distributed frequent subgraph mining method over MapReduce.
\alg\ generates a complete set of frequent subgraphs for a given minimum
support threshold. To ensure completeness, it constructs and retains all
patterns that have a non-zero support in the map phase of the mining, and
then in the reduce phase, it decides whether a pattern is frequent by
aggregating its support from different computing nodes.  To overcome the
dependency among the states of a mining process, \alg\ runs in an iterative
fashion, where the output from the reducers of iteration $i-1$ is used as an
input for the mappers in the iteration $i$. The mappers of iteration $i$
generate candidate subgraphs of size $i$~(number of edge), and also compute the
local support of the candidate pattern. The reducers of iteration $i$ then find
the true frequent subgraphs (of size $i$) by aggregating their local supports.
They also write the data in disk that are processed in subsequent iterations.

We claim the following contributions:

\begin{itemize}

 \item We introduce, \alg, a novel iterative MapReduce based frequent subgraph mining
algorithm, which is complete. 

 \item We design novel data structures to save and consequently propagate the
states of the mining process over different iterations.

 \item We empirically demonstrate the performance of \alg\ on synthetic as well
as real world large datasets.

\end{itemize}

\section{Related Works}\label{sec:RW}

There exist may algorithms for solving the in-memory version of frequent
subgraph mining task, most notable among them are
AGM~\cite{Inokuchi.Washio.ea:2000}, FSG~\cite{Kuramoche.Karypis:2001},
gSpan~\cite{Yan.Han:2002}, and Gaston~\cite{Nijssen.Kok:04}.  These methods
assume that the dataset is small and the mining task finishes in a reasonable
amount of time using an in-memory method.  To consider the large data scenario,
a few traditional database based graph mining algorithms, such as,
DB-Subdue~\cite{Sharma.Beera.ea:2004}, and DB-FSG~\cite{Chakravarthy.Subhesh:2008}
are also proposed. 

Researchers also considered shared memory parallel algorithms for frequent
subgraph mining. Cook et al.\ presented a parallel
version of their frequent subgraph mining algorithm
Subdue~\cite{Cook.Lawrence.ea:2001}. Wang et al.\ developed a parallel
toolkit~\cite{Wang.Parthasarathy:2004} for their
MotifMiner~\cite{Parthasarathy.Coatney:2002} algorithm. Meinl et al.\ created a
software named Parmol~\cite{Meinl.Worlein.ea:2006} which includes parallel
implementation of Mofa~\cite{Borgelt.Berthold:2002}, gSpan~\cite{Yan.Han:2002},
FFSG~\cite{Huan.Wang.ea:2003} and Gaston~\cite{Nijssen.Kok:04}.
ParSeMis\cite{parsemis:11} is another such tool that provides parallel
implementation of gSpan algorithm. To deal with the scalability problem caused
by the size of input graphs, there are couple of notable works,
PartMiner~\cite{Wang.Wynne.ea:2006} and
PartGraphMining~\cite{Nguyen.Orlowska.ea:2008}, which are based on the idea of
partitioning the graph data.

MapReduce framework has been used to mine frequent patterns that are simpler
than a graph, such as, a set~\cite{Chen.Yang.ea:12, Wang.Yang.ea:12,
Zhou.Zhong.ea:10, Li.Wang.ea:08}, and a sequence~\cite{Jeong.Choi.ea:12}. In
\cite{Hill.Srichandan.ea:2012}, the authors consider frequent subgraph mining
on MapReduce, however, their approach is inefficient due to various
shortcomings. The most notable is that in their method they do not adopt any
mechanism to avoid generating duplicate patterns. This cause an exponential
increase in the size of the candidate subgraph space; furthermore, the output
set contains duplicate copy of the same graph patterns that are hard to unify
as the user has to provide a subgraph isomorphism routine for this amendment.
Another problem with the above method is that it requires the user to specify
the number of MapReduce iterations.  Authors did not mention how to find the
total iteration count so that the algorithm is able to find all frequent
patterns for a given support.  One feasible way might be to set the iteration
count to the edge count of the largest transaction but that will be an
overkill of the resources. \alg\ does not suffer from any of the above
limitations.

There also exist a couple of works that mine subgraphs that are frequent considering
the embedding count of that graph in a single large
graph~\cite{Wu.Bai:2010}\cite{Liu.Jiang.ea:2009}.  Wu et al.\ developed a
distributed subgraph mining algorithm~\cite{Wu.Bai:2010} which requires the graph
diameter and the number of vertices for pattern matching. Liu et al.\ proposed an
algorithm MRPF~\cite{Liu.Jiang.ea:2009} that find motifs in prescription
compatibility network. Fatta et al.\ \cite{Fatta.Berthold:2006} used a search tree partitioning
approach, along with dynamic load balancing, for mining molecular structures. 

\section{Background}\label{sec:BG}
\begin{figure}
\begin{center}
    \includegraphics[width=0.4\textwidth]{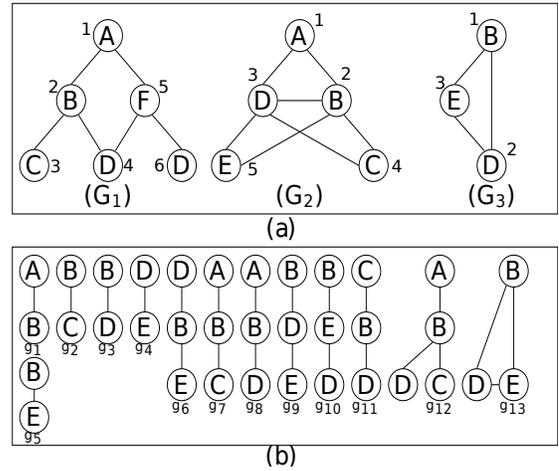}
\end{center}
  \caption{(a) Graph database with 3 graphs with labeled vertices (b) Frequent subgraph  of (a) with $minsup=2$}
  \label{fig:lattice_} 
\end{figure}

\subsection{Frequent Sub-graph Mining}

Let, ${\cal G} =\{G_1, G_2, \ldots, G_n \}$ be a graph database, where each
$G_i \in {\cal G}, \forall i=\{1 \ldots n\}$ represents a labeled, undirected
and connected graph. For a graph $g$, its size is defined as the number of
edges it contains. Now,  {\bf t}$(g) = \{G_i : g \subseteq G_i \in {\cal G}\},
\forall i = \{1 \ldots n \}$, is the {\em support-set} of the graph $g$  (here
the subset symbol denotes a subgraph relation). Thus, {\bf t}$(g)$ contains all the
graphs in ${\cal G}$ that has a subgraph isomorphic to $g$. The cardinality of
the {\em support-set} is called the {\em support} of $g$. $g$ is called
frequent if $support \geq \pi^{\textbf {min}}$, where $\pi^{\textbf {min}}$ is
predefined/user-specified \textit{minimum support (minsup)} threshold. The set
of frequent patterns are represented by ${\cal F}$. Based on the size (number
of edges) of a frequent pattern, we can partition ${\cal F}$ into a several
disjoint sets, ${\cal F}_i$ such that each of the ${\cal F}_i$ contains
frequent patterns of size $i$ only.

\textbf{Example:} Figure \ref{fig:lattice_}(a) shows a database with 3
vertex labeled graphs ($G_1$, $G_2$ and $G_3$). With $\pi^{\textbf {min}}=2$, there are
thirteen frequent subgraphs as shown in Figure~\ref{fig:lattice_}(b).
Note that, For the sack of simplicity in this example we assume that the edges of the input graphs are unlabeled. 
But \alg\ is designed and developed to handle labels both on vertices and edges.~$\qed$

\subsection{MapReduce}

MapReduce is a programming model that enables distributed computation over
massive data~\cite{Dean.Ghemawat:2008}. The model provides two abstract
functions: {\em map}, and {\em reduce}. Map corresponds to the ``map'' function
and reduce corresponds to the ``fold'' function in functional programming.
Based on its role, a worker node in MapReduce is called a mapper or a reducer.
A mapper takes a collection of  (key, value) pairs and applies the map function
on each of the pairs to generate an arbitrary number of (key,value) pairs as
intermediate output. The reducer aggregates all the values that have the same
key in a sorted list, and applies the reduce function on that list. It also
writes the output to the output file. The files (input and output) of MapReduce
are managed by a distributed file system. Hadoop is an open-source
implementation of MapReduce programming model written in Java language.

\begin{figure} [!ht]
  \fbox{
  \begin{minipage}{7 cm}
      \begin{flushleft}
        {\bf Iterative\_MapReduce}(): \\
        1.~~{\bf While}(Condition)\\
	2.~~~~Execute MapReduce Job\\
	3.~~~~Write result to DFS\\
	4.~~~~Update condition\\
    \end{flushleft}
\end{minipage}
}
  \caption{Iterative MapReduce Algorithm}
   \label{fig:iterrative_alg}
\end{figure}
\subsubsection{Iterative MapReduce}

Iterative MapReduce~\cite{Lin.Dyer:10} can be defined as a multi staged
execution of map and reduce function pair in a cyclic fashion, i.e. the output of
the stage $i$ reducers is used as an input of the stage $i+1$ mappers. An
external condition decides the termination of the job.  Pseudo code for
iterative MapReduce algorithm is presented in Figure~\ref{fig:iterrative_alg}.

\section{Method}\label{sec:ALG}

\alg\ is designed as an iterative MapReduce process. At the beginning of
iteration $i$, \alg\ has at its disposal all the frequent patterns of size
$i-1$ (${\cal F}_{i-1}$), and at the end of iteration $i$, it returns all the
frequent patterns of size $i$, $({\cal F}_i)$. Note that, in this work, the size of
a graph is equal to the number of edges it contains. Now, for a mining task,
if ${\cal F}$ is the set of frequent patterns \alg\ runs for a total of $l$
iterations, where $l$ is equal to the size of the largest graph in ${\cal F}$. 

To distribute a frequent subgraph mining (FSM) task, \alg\ partitions the graph
dataset ${\cal G}= \{G_i\}_{i=1 \ldots n}$ into $k$ disjoint partitions, such
that each partition contains roughly equal number of graphs; thus it mainly
distributes the support counting (discussed in details later) subroutine of a
frequent pattern mining algorithm. Conceptually, each node of \alg\ runs an
independent FSM task over a graph dataset which is $1/k$'th of the size of
$|{\cal G}|$. The FSM algorithm that \alg\ implements is an adaptation of the
baseline algorithm shown in Figure~\ref{fig:baseAlg}, which runs in a
sequential machine. Below we provides more details of this algorithm.

\begin{figure} [!ht]
  \fbox{
  \begin{minipage}{7 cm}
      \begin{flushleft}
        // ${\cal G}$ is the database \\
        // $k$ is initialize to 1 \\
        {\bf Mining\_Frequent\_Subgraph}(${\cal G}, minsup$): \\
        0.~~{\bf Populate} ${\cal F}_1$\\
        1.~~{\bf while} ${\cal F}_k \ne \emptyset$ \\
        2.~~~~${\cal C}_{k+1}$ = {\bf Candidate\_generation}(${\cal F}_k, {\cal G}$) \\
        2.~~~~{\bf forall} $c \in {\cal C}_{k+1}$\\
        3.~~~~~~\textbf{if} {\bf isomorphism\_checking($c$)} = {\bf true}\\
        4.~~~~~~~~{\bf{support\_counting}($c, {\cal G}$)}\\
        5.~~~~~~~~{\bf if} $c.sup \geq minsup$\\
        6.~~~~~~~~~~${\cal F}_{k+1}$ = ${\cal F}_{k+1} \bigcup \{c\}$\\
        7.~~~~$k=k+1$\\
        8.~~{\bf return} $\bigcup_{i=1\cdots k-1}{{\cal F}_i}$\\
    \end{flushleft}
\end{minipage}
}
  \caption{Frequent subgraph mining algorithm with breadth-first candidate enumeration}
   \label{fig:baseAlg}
\end{figure}

\subsection{Baseline frequent subgraph mining algorithm} \label{sec:Baseline}

The pseudo-code shown in Figure~\ref{fig:baseAlg} implements an FSM algorithm
that follows a typical candidate-generation-and-test paradigm with
breadth-first candidate enumeration. In this paradigm, the mining task starts
with frequent patterns of size one (single edge patterns), denoted as ${\cal
F}_1$ (Line 0). Then in each of the iterations of the while loop (Line 1-6),
the method progressively finds ${\cal F}_2, {\cal F}_3$ and so on until the
entire frequent pattern set (${\cal F}$) is obtained. If ${\cal F}_k$ is
non-empty at the end of an iteration of the above while loop, from each of the
frequent patterns in ${\cal F}_k$ the mining method creates possible candidate
frequent patterns of size $k$+1 (Line 2). These candidate patterns are
represented as the set ${\cal C}$. For each of the candidate patterns, the
mining method computes the pattern's support against the dataset ${\cal G}$
(Line 5). If the support is higher than the minimum support threshold
($minsup$), the given pattern is frequent, and is stored in the set ${\cal
F}_{k+1}$ (Line 6).  Before support counting, the method also ensures that
different isomorphic forms of a unique candidate patterns are unified and only
one such copy is processed by the algorithm (Line 4). Once all the frequent
patterns of size $k+1$ are obtained, the while loop in Line 1 to 7 continues.
Thus each iteration of the while loop obtains the set of frequent patterns of a
fixed size, and the process continues until all the frequent patterns are
obtained. In Line 8, the FSM algorithm returns the union of ${\cal F}_i:1\le i
\le k-1$.

Below, we provide a short description of each of the subroutines that are
called in the pseudo-code.

\begin{figure}
\begin{center}
    \includegraphics[width=0.5\textwidth]{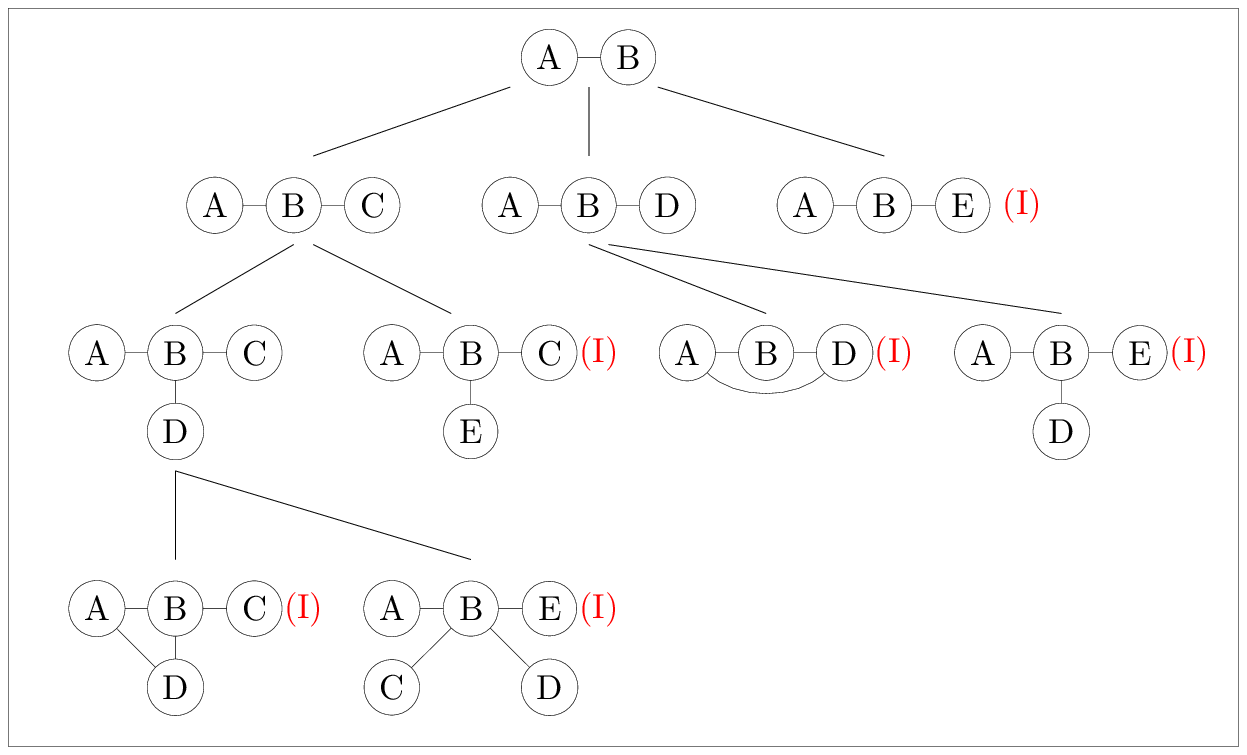}
\end{center}
  \caption{Candidate generation subtree rooted under $A-B$}
\vspace{-0.2in}
  \label{fig:baseline_candgen} 
\end{figure}

\subsubsection{Candidate Generation}\label{sec:Candgen}

Given a frequent pattern (say, $c$) of size $k$, this step adjoins a frequent
edge (which belongs to ${\cal F}_1$) with $c$ to obtain a candidate pattern $d$
of size $k+1$. If $d$ contains an additional vertex then the added edge is
called a {\em forward edge}, otherwise it is called a {\em back edge}; the
latter simply connects two of the existing vertices of $c$. Additional vertex
of a forward edge is given an integer id, which is the largest integer id
following the ids of the existing vertices of $c$; thus the vertex-id stands for the
order in which the forward edges are adjoined while building a candidate pattern.  In
graph mining terminology, $c$ is called the parent of $d$, and $d$ is a child
of $c$, and based on this parent-child relationship we can arrange the set of
candidate patterns of a mining task in a candidate generation tree (see
Figure~\ref{fig:baseline_candgen}). Note that, if $d$ has $k+1$ edges, based on
the order how its edges has been adjoined, $d$ could have many different
generation paths in a candidate generation tree; however, in all FSM
algorithms, only one of the generation paths is considered valid, so that
multiple copies of a candidate pattern is not generated. With this restriction
of candidate generation, the candidate generation tree of an FSM task can be
unambiguously defined.

Existing FSM algorithms also impose restriction on the extension nodes of the
parent pattern so that redundant generation paths can be reduced.  One such
restriction that is used in the popular gSpan algorithm~\cite{Yan.Han:2002} is
called rightmost path generation that allows adjoining edges only with vertices
on the rightmost path.  Simple put, ``right most vertex''~(RMV) is the vertex
with the largest id in a candidate subgraph and ``right most path''~(RMP) is
the shortest path from the lowest id vertex to the RMV strictly following
forward edges.

\textbf{Example:} In Figure~\ref{fig:baseline_candgen}, we show a part of the
candidate generation tree of the FSM task that we define in
Figure~\ref{fig:lattice_}. Suppose we have explored all frequent level-1
patterns~($g_1$ to $g_4$ in Figure~\ref{fig:lattice_}(b)) and we want to
generate candidates from one of those patterns, namely $A-B$.
Figure~\ref{fig:baseline_candgen} shows the part of the candidate generation
subtree that is rooted at the pattern $A-B$.  The nodes at the level 2 (root of the
subtree is level 1) of this subtree shows all possible candidate patterns of
size 2 that are built by adjoining the edge $B-C$, $B-D$, and $B-E$,
respectively, with the pattern $A-B$. Note that, all these candidate patterns
are created by introducing a forward edge as adding a back edge will create a
multi-graph, which we do not allow. Also note, we do not extend $A-B$ with
another copy of $A-B$ to create the pattern $A-B-A$ because none of the
database graphs in Figure~\ref{fig:lattice_}(a) has multiple copies of the edge
$A-B$. Among these three candidate patterns, the pattern $A-B-E$ is infrequent
which is denoted with the mark $(I)$ near the pattern. The remaining two
patterns are frequent and are extended further to generate level-3 candidates
For example, the pattern $A-B-D$ is extended with an back-edge to obtain the
triangle pattern $A-B-D$ (all level-3 or level-4 nodes are not shown).~$\qed$

There are other important observations in Figure~\ref{fig:baseline_candgen}.
First, the extension of a pattern is only made on the rightmost path of that
pattern. For visual clarification, for each pattern we draw its rightmost path
along a horizontal line.  The second observation is that, the duplicate
generation paths are avoided; for example, the pattern $B-\{A,C,D\}$
(the first pattern from the left on level-3) is generated from the pattern
$A-B-C$, but not from the pattern $A-B-D$,

\begin{figure}
\begin{center}
    \includegraphics[width=0.3\textwidth]{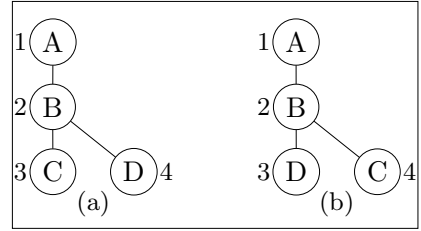}
\end{center}
  \caption{Graph isomorphism}
\vspace{-0.2in}
  \label{fig:baseline_isotest} 
\end{figure}

\subsubsection{Isomorphism checking}\label{sec:isomorphism_check}

As we mention in previous paragraph, a candidate pattern can be generated from
multiple generation paths, but only one such path is explored during the candidate
generation step and the remaining paths are identified and subsequently
ignored. To identify invalid candidate generation paths, a graph mining
algorithm needs to solve the graph isomorphism task, as the duplicate copies of
a candidate patterns are isomorphic to each other. A well-known method for
identifying graph isomorphism is to use canonical coding scheme, which
serializes the edges of a graph using a prescribed order and generates a string
such that all isomorphic graphs will generate the same string. There are many
different canonical coding schemes, min-dfs-code is one of those which is
used in~\cite{Yan.Han:2002}. According to this scheme, the generation path 
of a pattern in which the insertion order of the edges matches with the
edge ordering in the min-dfs-code is considered as the valid generation path,
and the remaining generation paths are considered as duplicate and hence ignored.
\alg\ uses min-dfs-code based canonical coding for isomorphism checking.

\textbf{Example:} Figure~\ref{fig:baseline_isotest}(a)
and Figure~\ref{fig:baseline_isotest}(b) shows two isomorphic forms of the 
pattern $B-\{A,C,D\}$, however during candidate generation phase the
first is generated from $A-B-C$ wheres the second would have been generated from 
$A-B-D$. According to the canonical coding scheme in~\cite{Yan.Han:2002}, the
pattern in Figure~\ref{fig:baseline_isotest}(a) is mapped to a code-string
$(1,2,A,B)(2,3,B,C)(2,4,B,D)$ in which each parenthesized part is an edge written
in the format $(id_1, id_2, label_1, label_2)$. Using the same coding scheme,
the pattern in Figure~\ref{fig:baseline_isotest}(b) is mapped to the string
$(1,2,A,B)(2,3,B,D)(2,4,B,C)$. However, the min-dfs-code of the pattern 
$B-\{A,C,D\}$ is $(1,2,A,B)(2,3,B,C)(2,4,B,D)$, which matches with the isomorphic
form shown in Figure~\ref{fig:baseline_isotest}(a); thus the pattern will only be generated by
extending $A-B-C$. Other generation paths, including the one that extends $A-B-D$
are invalid and hence are ignored after performing isomorphism checking.~$\qed$

\begin{figure}
\begin{center}
    \includegraphics[width=0.25\textwidth]{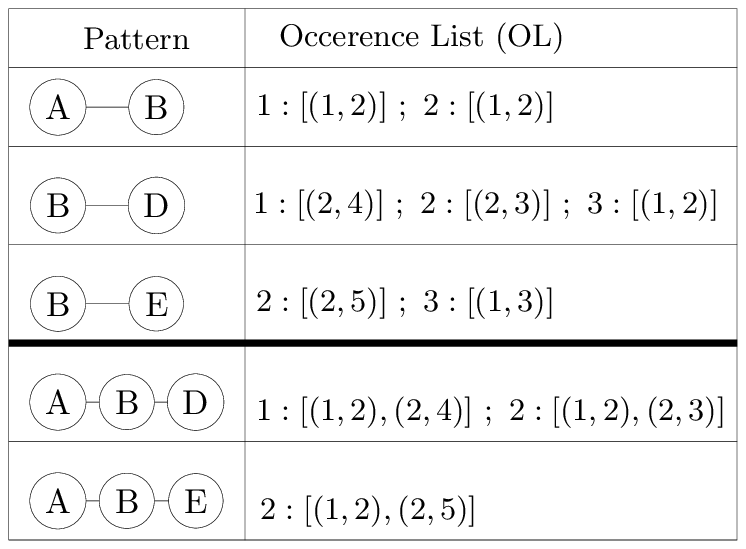}
\end{center}
  \caption{ Support Counting}
\vspace{-0.2in}
  \label{fig:baseline_support} 
\end{figure}

\subsubsection{Support Counting}

Support counting of a graph pattern $g$ is important to determine whether
$g$ is frequent or not.  To count $g$'s support we need to find the
database graphs in which $g$ is embedded. This mechanism requires solving a
subgraph isomorphism problem, which is ${\cal NP}$-complete. One possible way
to compute the support of a pattern without explicitly performing the subgraph isomorphism
test across all database graphs is to maintain the occurrence-list (OL)
of a pattern; such a list stores the embedding of the pattern (in terms of
vertex id) in each of the database graphs where the pattern exists. When a
pattern is extended to obtain a child pattern in the candidate generation step,
the embedding of the child pattern must include the embedding of the parent
pattern, thus the occurrence-list of the child pattern can be generated
efficiently from the occurrence list of its parent. Then the support of a child
pattern can be obtained trivially from its occurrence-list.

\textbf{Example:} In Figure~\ref{fig:baseline_support} we illustrate a simple
scenario of support counting based on occurrence list. In the top three rows of
Figure~\ref{fig:baseline_support}, we show the OL of the pattern $A-B$, $B-D$
and $B-E$. The Pattern $A-B$ occurs in Graph $G_1$ and $G_2$ in vertex pair
$(1,2)$ and $(1,2)$, respectively; so its OL is: 1:[(1,2)]; 2:[(1,2)].  If we
adjoin $B-D$ with the pattern $A-B$ and form the pattern $A-B-D$, then we can
construct the OL of $A-B-D$ (shown in 4th row) by intersecting the OLs of $A-B$ and $B-D$. Note
that, the intersection considers both the graph ids and the vertex ids in the OLs.
By counting the graph ids present in an OL we can compute the support of that pattern. In 
Figure~\ref{fig:baseline_support}, the pattern $A-B-D$ is frequent given minimum
support $2$ but the pattern $A-B-E$ is not frequent.~$\qed$

\begin{figure} [!t]
  \fbox{
  \begin{minipage}{7 cm}
      \begin{flushleft}
        {\bf Mapper\_FSG}(${\cal F}_k^p \langle x.\textrm{min-dfs-code}, x.obj \rangle$): \\
        1.~~${\cal C}_{k+1}$ = {\bf Candidate\_generation}(${\cal F}_k^p$) \\
        2.~~{\bf forall} $c \in {\cal C}_{k+1}$\\
        3.~~~~\textbf{if} {\bf isomorphism\_checking($c$)} = {\bf true}\\
        4.~~~~~~{\bf{populate\_occurrence\_List}($c$)}\\
        5.~~~~~~{\bf if} $length(c.occurrence\_List) > 0$\\
        6.~~~~~~~~{\bf emit} $(c.$min-dfs-code$~,~c.obj)$
    \end{flushleft}
\end{minipage}
}
  \caption{Mapper of Distributed Frequent Subgraph Mining Algorithm}
  \vspace{-0.1in}
   \label{fig:Distributed_alg_mapper}
\end{figure}

\begin{figure} []
  \fbox{
  \begin{minipage}{7.2 cm}
      \begin{flushleft}
        {\bf Reducer\_FSG}($c.$min-dfs-code$~,\langle~c.obj~\rangle$): \\
        1.~~{\bf forall} $obj \in \langle~c.obj~\rangle$\\
        2.~~~~support += $length(obj.OL)$\\
        3.~~\textbf{if} {support $\geq$ \textit{minsup}}\\
        4.~~~~{\bf forall} $obj \in \langle~c.obj~\rangle$\\
        5.~~~~~~{\bf write} $(c.$min-dfs-code$~,~obj)$ {\bf to HDFS}\\
    \end{flushleft}
\end{minipage}
}
  \caption{Reducer of Distributed Frequent Subgraph Mining Algorithm}
  \vspace{-0.2in}
   \label{fig:Distributed_alg_reducer}
\end{figure}

\subsection{Distributed paradigm of \alg}

An important observation regarding the baseline FSM
algorithm~(Figure~\ref{fig:baseAlg}) is that it obtains all the frequent
patterns of size $k$ in one iteration of while loop from Line 1 to Line 6. The
tasks in such an iteration comprise to one MapReduce iteration of \alg. Another
observation is that, when the FSM algorithm generates the candidates of size
$k+1$, it requires the frequent patterns of size $k$ (${\cal F}_k$). In an
iterative MapReduce, there is no communication between subsequent iterations.
So, $k+1$'th iteration of \alg\ obtains ${\cal F}_k$ from the disk which is
written by the reducers at the end of the $k$'th iteration. A final observation
is that, deciding whether a given pattern is frequent or not requires counting
it's support over all the graphs in the dataset (${\cal G}$).  However, as we
mentioned earlier each node of \alg\ works only on a disjoint partition of
${\cal G}$. So, \alg\ requires to aggregate the local support from each node
to perform the task in Line 5.  From the above observations we identify the
distribution of mining task of \alg\ among the mappers and the reducers. 

Figure~\ref{fig:Distributed_alg_mapper} shows the pseudo-code of a mapper. The
argument ${\cal F}_k^p$ represents the set of size-$k$ frequent subgraphs having 
non-zero support in a specific partition $p$. The mapper reads it from Hadoop
Distributed File Systems (HDFS). Each pattern (say $x$) in ${\cal F}_k^p$ is read as a key-value
pair. The key is the min-dfs-code of the pattern ($x$.min-dfs-code) and the
value is a pattern object~($x.obj$), here ``object'' stands for its usual
meaning from the object oriented programming. This pattern object contains all
the necessary information of a pattern i.e., its support, neighborhood lists,
and occurrence list. It also contains
additional data structure that are used for facilitating candidate generation from this pattern.
We will discuss the pattern object in details in a later section. The mapper then
generates all possible candidates of size $k+1$~(Line 1) by extending each of the patterns in ${\cal F}_k^p$.
For each of the generated candidates (say, $c$), the mapper performs isomorphism checking
to confirm whether $c$ is generated from a valid generation path;
in other words, it tests whether $c$ passes the min-dfs-code based
isomorphism test~(Line 3).  For successful candidates, the mapper populates
their occurrence list~(Line 4) over the database graphs in the 
partition $p$. If the occurrence list of a candidate pattern is
non-empty, the mapper constructs a key-value pair, such as, $(c.\textrm{min-dfs-code}, c.obj)$
and emits the constructed pair for the reducers to receive (Line 6).

Figure~\ref{fig:Distributed_alg_reducer} shows the pseudo code for a reducer in
distributed frequent subgraph mining. The reducer receives a set of key-value pairs, where
the key is the min-dfs-code of a pattern namely $c$.min-dfs-code and the value
is a list of $c.obj$'s constructed from all partitions where the pattern $c$
has a non-zero support.  Reducer than iterates~(Line 1) over every $c.obj$ and
from the length of the occurrence list of each $c.obj$ it computes the
aggregated support of $c$.  If the aggregated support is equal or higher than the
minimum support threshold (Line 3), the reducer writes each element in the list
paired with the min-dfs-code of $c$ in HDFS for the next iteration mappers.

\begin{figure}
\begin{center}
    \includegraphics[width=0.5\textwidth]{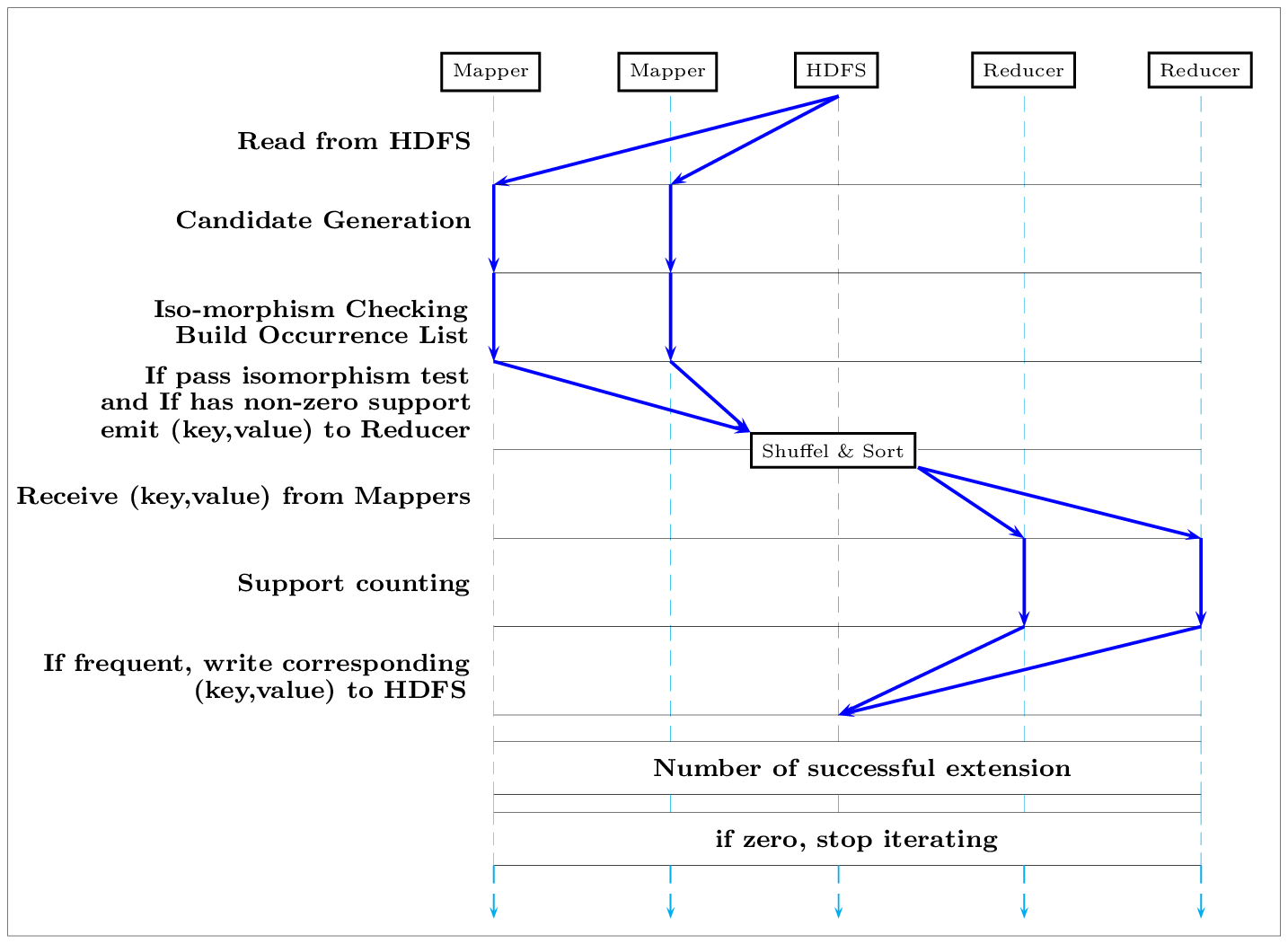}
\end{center}
  \caption{ Execution Flow of \alg\ }
\vspace{-0.2in}
  \label{fig:execution_flow} 
\end{figure}

Figure~\ref{fig:execution_flow} illustrates the execution flow of \alg\ . The
execution starts from the mappers as they read the key-value pair of size $k$
patterns from the HDFS. As shown in the Figure~\ref{fig:execution_flow}, the mappers generate all
possible $k+1$-size candidate patterns and perform the isomorphism checking.
For a pattern of size $k+1$ that passes the isomorphism test and has a non-zero occurrence,
the mapper builds its key-value pair and emits that for the reducers.
These key-value pairs are shuffled and sorted by the key field and each reducer
receives a list of values with the same key field. The reducers then compute the support of
the candidate patterns. If a pattern is frequent, the reducer writes appropriate key-value
pairs in the HDFS for the mappers of the next iteration. If the number of
frequent $k+1$ size pattern is zero, execution of \alg\ halts.

\begin{figure}
\begin{center}
    \includegraphics[width=2.5in,height=3in]{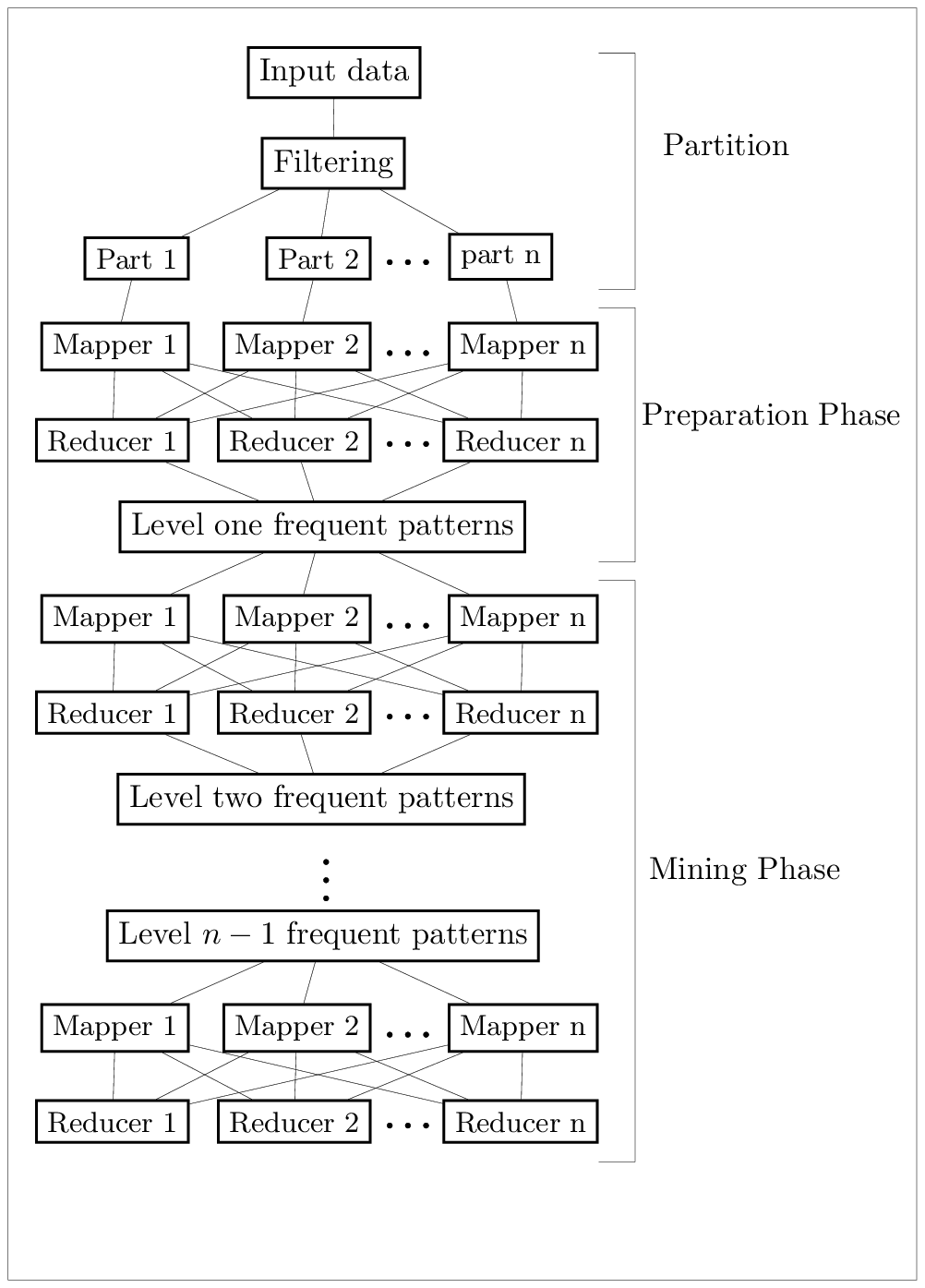}
\end{center}
  \caption{(a) Framework of \alg\ }
\vspace{-0.2in}
  \label{fig:framework} 
\end{figure}

\subsection{Framework of \alg\ }\label{sec:framework}

\alg\ has three important phases: data partition, preparation phase and mining
phase. In data partition phase \alg\ creates the partitions of input data along
with the omission of infrequent edges from the input graphs. Preparation and
mining phase performs the actual mining task. Figure~\ref{fig:framework} shows
a flow diagram of different phases for a frequent subgraph mining task using
\alg. 

Below, we present an in-depth discussion of each of the phases.

\begin{figure}
\begin{center}
    \includegraphics[width=0.4\textwidth]{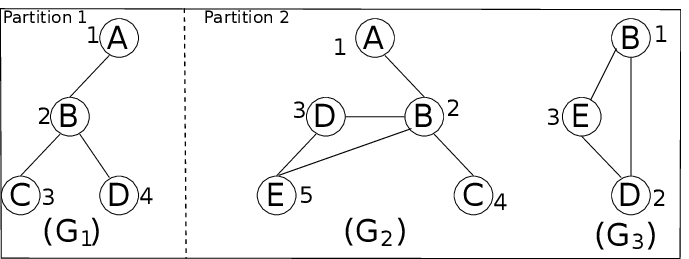}
\end{center}
  \caption{Input data after partition and filtering phase}
\vspace{-0.2in}
  \label{fig:partition_dataset} 
\end{figure}

\subsubsection{Data partition}\label{sec:partition}

In data partition phase, \alg\ splits the input graph dataset ($\cal G$)
into many partitions. One straightforward partition scheme is to distribute the
graphs so that each partition contains the same number of graphs from ${\cal G}$. This works
well for most of the datasets. However, for datasets where the size (edge
count) of the graphs in a dataset vary substantially, \alg\ offers another
splitting option in which the total number of edges aggregated over the graphs
in a partition, are close to each other. In experiment section, we show that
the latter partition scheme has a better runtime performance as it improves
the load balancing factor of a MapReduce job. For \alg, the number of
partition is also an important tuning parameter. In experiment section, we also
show that for achieving optimal performance, the number of partitions for \alg\
should be substantially larger than the number of partition in a typical
MapReduce task. 

During the partition phase, input dataset also goes through a filtering
procedure that removes the infrequent edges from all the input graphs. While
reading the graph database for partitioning, \alg\ computes the
support-list of each of the edges from which it identifies the edges
that are infrequent for the given minimum support threshold.

\textbf{Example}: For the graph dataset in Figure~\ref{fig:lattice_}, for a
minimum support threshold of 2, the edges $A-B, B-C, B-D$, $D-E$ and $B-E$ are
frequent and the remaining edges are infrequent. Now suppose \alg\ makes two
partitions for this dataset such that the first partition contains $G_1$, and
the second partition contains $G_2$ and $G_3$. While making these partitions
\alg\ filters the infrequent edges. Figure~\ref{fig:partition_dataset} shows
the partitioning where the infrequent edges are stripped off from the database
graphs.~$\qed$

\begin{figure}
\begin{center}
    \includegraphics[width=0.5\textwidth]{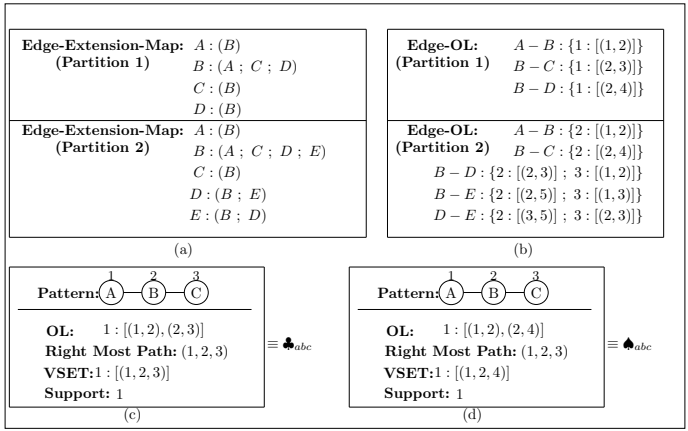}
\end{center}
  \caption{The static data structures and $A-B-C$ pattern object in partition 1 and 2 (a) Edge-extension Map (b) Edge-OL (c) and (d) $A-B-C$ Pattern object}
\vspace{-0.2in}
  \label{fig:noniterative_phase} 
\end{figure}

\subsubsection{Preparation Phase}\label{sec:noniter}

The mappers in this phase prepare some partition specific data structures such
that for each partition there is a distinct copy of these data structures. They
are static for a partition in the sense that they are same for all patterns
generated from a partition. The first of such data structure is called {\em
edge-extension-map}, which is used for any candidate generation that happens
over the entire mining session. It stores the possible extension from a vertex
considering the edges that exists in the graphs of a partition. For example,
the graphs in partition two have edges such as $B-D, B-C, B-A$, and $B-E$. So,
while generating candidates, if $B$ is an extension stub, the vertex $A, C, D$
or $E$ can be the possible vertex label of the vertex that is at the opposite
end of an adjoined edge. This information is stored in the Edge-extension-map
data structure for each of the vertex label that exist in a partition.  The
second data structure is called {\em edge-OL}, it stores the occurrence list of
each of the edges that exist in a partition; \alg\ use it for counting the support of
a candidate pattern which is done by intersecting the OL of a parent pattern with the OL of
an adjoint edge. 

\textbf{Example}: Figure~\ref{fig:noniterative_phase} (a) and
Figure~\ref{fig:noniterative_phase} (b) shows these data structures for the
Partition 1 and 2 defined in Figure~\ref{fig:partition_dataset}.  In partition
1, the edge-extension choice from a vertex with label $D$ is only $B$ (shown as
$D:(B)$), as in this partition $B-D$ is the only frequent edge with a $B$
vertex. On the other hand, the corresponding choice for partition 2 is $B$ and
$E$ (shown as, $D:(B;E)$), because in partition 2 we have two edges, namely $B-D$
and $D-E$ that involve the $D$ vertex. In partition 1, the edge $B-D$ occurs
in $G_1$ at vertex id $(2, 4)$; on the other hand in partition 2, the same edge
occurs in $G_2$ and $G_3$ at vertex id $(2, 3)$ and $(1,2)$, respectively. These
information are encoded in the {\em edge-OL} data structures of these partitions
as shown in this figure.~$\qed$

The mappers in the preparation phase also starts the mining task by emitting the
frequent single edge pattern as key-value pair. Note that, since the partition
phase have filtered out all the infrequent edges, all single edges that exist
in any graph of any partition is frequent. As we mentioned earlier the key of a
pattern is its min-dfs-code and the value is the pattern object.  Each pattern
object have four essential attributes: (a) Occurrence List~(OL) that stores the 
embedding of the pattern in each graph in the partition, (b) \textit{Right-Most-Path}
(c) VSET that  stores the embedding of the \textit{Right Most Path} in each graph in 
the partition, and (d) support value. Mappers in the preparation phase computes
the min-dfs-code and creates the pattern object for each single-edge patterns.

While emitting the key-value pair to a reducer, the mappers also bundle the
static data structures, edge-extension-map and edge-OL with each of the pattern
object.  This is an wasteful overhead in terms of network traffic and disk
space, as the static data structure is common for all the patterns that are
ever generated from a partition. We had to adopt this strategy, because to the
best of our knowledge there is no provision in Hadoop to keep a global data
structure that can be accessed by each of the mappers. \alg\ uses Java serialization to 
convert these objects in to byte stream while sending them as value in a key-value pair. 

The reducers of this phase actually do nothing but writing the input key-value pairs in HDFS since 
all the single length patterns that the mappers send are frequent. 
In Figure~\ref{fig:framework}, the second block portrays the preparation phase.

\textbf{Example}: Figure~\ref{fig:noniterative_phase}(c) and ~\ref{fig:noniterative_phase}(d)
exhibits the \textit{Pattern} object along with their attributes for the
pattern $A-B-C$ in partition 1 and 2, respectively. The attribute OL records
the occurrence of this pattern in the corresponding database graphs; if a
pattern has multiple embeddings in a database graph all such embeddings are
stored. Right-Most-Path records the id of the right-most-path vertices in the
pattern object and VSET stores the corresponding ids in the database graphs.
Like OL, VSET is also a set and it stores information for multiple embedding if
it applies. Finally, Support stores the support value of the pattern.  In the
following discussion, we use $\clubsuit$ and $\spadesuit$ to denote a pattern
object from partition 1~($G_1$) and 2~($G_2,G_3$), respectively. For example,
$\spadesuit_{abc}$ identifies the pattern $A-B-C$ from partition
2 as shown in Figure~\ref{fig:noniterative_phase}(d).~$\qed$

\begin{figure}
\begin{center}
    \includegraphics[width=0.5\textwidth]{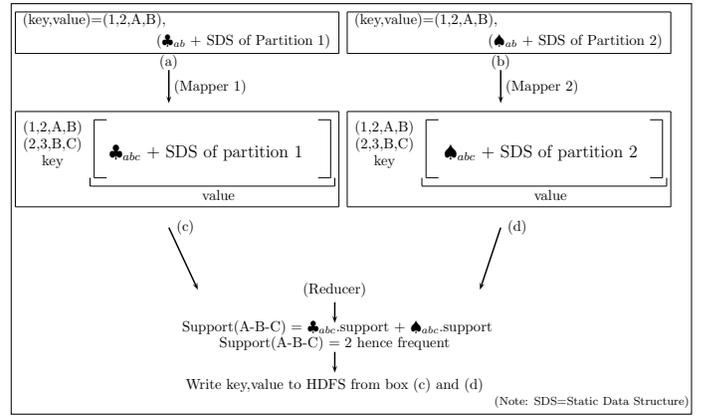}
\end{center}
  \caption{One iteration of the mining phase of \alg\ with respect to pattern A-B}
\vspace{-0.2in}
  \label{fig:iterative_phase} 
\end{figure}

\subsubsection{Mining Phase}

In this phase, mining process discovers all possible frequent subgraphs through 
iteration. Preparation phase populates all frequent subgraphs of size one and 
writes it in the distributed file system. Iterative job starts by reading these 
from HDFS. Each of the mappers of an ongoing iteration is responsible for performing 
the mining task over a particular chunk of the data written in HDFS by the preparation phase. 
The map function of mining phase reconstructs all the static data structures that
are required to generate candidate patterns from the current pattern. Using the static
data structures and the pattern object, the mappers can independently execute
the subroutine that is shown in Figure~\ref{fig:Distributed_alg_mapper}. The
reducers in this phase simply execute the routine in Figure~\ref{fig:Distributed_alg_reducer}.

{\bf Example}:
Figure~\ref{fig:iterative_phase} provides a detail walk-through of a MapReduce
job in iteration phase with respect to the partitioned toy dataset mentioned in
Figure~\ref{fig:partition_dataset}.  Figure~\ref{fig:iterative_phase}(a) and
(b) indicates the key-value pairs for pattern $A-B$ from partition $1$ and $2$.
Suppose these key-value pairs are feed as input for Mapper 1 and Mapper 2 in
Figure~\ref{fig:iterative_phase}.  The mappers first extract all data
structures from the value field of key-value pair including the
partition-specific static data structures such as edge-extension-map and
edge-OL. Then they perform the steps mentioned in
Figure~\ref{fig:Distributed_alg_mapper}.  Figure~\ref{fig:iterative_phase}(c)
and (d) show the key-value pairs of $A-B-C$ that are generated by Mapper
1 and Mapper 2 by extending the pattern $A-B$. The Reducer 
collects all the values for a key and computes the support of the pattern of
of the given key by adding the supports from
individual partitions. In this example, the support of the pattern $A-B-C$ is $2$;
since $minsup$=2, this pattern is frequent. Reducer then writes the key-value pairs corresponding to the
pattern $A-B-C$ in HDFS.~$\qed$

\subsection{Pseudo Code}
In this section, we present the pseudo code of each phase using the syntax of Hadoop framework. 
Figure~\ref{fig:partition}, \ref{fig:noniteralg} and
\ref{fig:iteralg} demonstrate data partitioning, preparation and mining phase, respectively.
In Line 3 and 4 of Figure~\ref{fig:partition}, \alg\ performs the filtering and partitioning of the 
input dataset and writes each partition and write in 
the HDFS. In line 1-7 of Figure~\ref{fig:noniteralg} the mappers generates static data structure
along with emit key-value pair of all single length pattern to reducer.
Since all patterns are frequent, Reducers just relay
the input key,value pair to the output file in HDFS. In Line 1-3 of Mapper\_mining function 
in Figure~\ref{fig:iteralg}, the mappers reconstruct the pattern object of size $k$ along with
the static data structures and generates the candidates from the current pattern.
In Line 4-9, the mappers iterate over all possible candidates of size $k+1$  and on success
in isomorphism and occurrence list test mappers emits the key-value pair for the reducer
Reducer\_mining function computes the aggregate support~(Line 2) of the pattern
and if the pattern is frequent, they write back the key,value pairs
to HDFS for the mappers of the next iteration.

\begin{figure} [!ht]
  \scriptsize
  \fbox{
  \begin{minipage}{7 cm}
      \begin{flushleft}
        {\bf Partition\_data}(${\cal D}$): \\
        1.~~create a data directory in distributed file system\\
        2.~~{\bf while} data available in ${\cal D}$\\ 
        3.~~~~{\bf $partition$ = create\_partition()}\\
        4.~~~~write $partition$ to file $partition_i$ in data directory\\
    \end{flushleft}
\end{minipage}
}
  \caption{Data Partitioning Phase}
  \vspace{-0.2in}
   \label{fig:partition}
\end{figure}

\begin{figure} [!ht]
  \scriptsize
  \fbox{
  \begin{minipage}{7 cm}
      \begin{flushleft}
        //~key~~~~=~offset \\
        //~value~~=~location of partition file in data directory \\
        {\bf Mapper\_preparation}(Long $key$, Text $value$): \\
        1.~~{\bf Generate\_Level\_one\_OL($value$)}\\
        2.~~{\bf Generate\_Level\_one\_MAP($value$)}\\
        3.~~{\bf $P$ = get\_single\_length\_patterns()}\\
        4.~~{\bf forall $P_i$ in $P$}:\\
        5.~~~~{\bf $intermediate\_key$ = min-dfs-code($P_i$)}\\
        6.~~~~{\bf $intermediate\_value$ = serialize($P_i$)}\\
        7.~~~~{\bf emit($intermediate\_key$,$intermediate\_value$)}\\
        \vspace{3mm}
        // key~~~~~= min-dfs-code \\
        // values~~= List of Byte-stream of a pattern object in all partitions\\
        {\bf Reducer\_preparation}(Text $key$, BytesWritable $\langle$ values $\rangle$): \\
        1.~~~~{\bf for all $value$ in $values$:}\\
        2.~~~~~~{\bf write\_to\_file($key$,$value$)}\\ 
    \end{flushleft}
\end{minipage}
}
  \caption{Preparation Phase }
  \vspace{-0.2in}
   \label{fig:noniteralg}
\end{figure}

\begin{figure} [!ht]
  \scriptsize
  \fbox{
  \begin{minipage}{7 cm}
      \begin{flushleft}
        //~key~~~~=~min-dfs-code \\
        //~value~~=~Byte-stream of pattern object for iteration i-1 \\
        {\bf Mapper\_mining}(Long $key$, BytesWritable $value$): \\
        1.~~{\bf $p$~=~reconstruct\_pattern($value$)}\\
        2.~~{\bf reconstruct\_all\_data-structures($value$)}\\
        3.~~{\bf $P$~=~Candidate\_generation($p$)}\\
        4.~~{\bf forall} $P_i$ in $P$:\\
        5.~~~~{\bf if pass\_isomorphism\_test($P_i$) = true}\\
        6.~~~~~~{\bf if $length(P_i.OL) > 0$}\\
        7.~~~~~~~~{\bf $intermediate\_key$ = min-dfs-code($P_i$)}\\
        8.~~~~~~~~{\bf $intermediate\_value$ = serialize($P_i$)}\\
        9.~~~~~~~~{\bf emit($intermediate\_key$,$intermediate\_value$)}\\
        \vspace{3mm}
        // key~~~~~= min-dfs-code \\
        // values~~= List of Byte-stream of a pattern object in all partitions\\
        {\bf Reducer\_mining}(Text $key$, BytesWritable $\langle values \rangle$): \\
        1.~~{\bf for all $value$ in $values$}:\\
        2.~~~~$support$ += {\bf get\_support($value$)}\\
        3.~~{\bf if} $support~\geq~minimum\_support$\\
        4.~~~~{\bf for all $value$ in $values$}:\\
        5.~~~~~~\bf{write\_to\_file($key,value$)}\\ 
    \end{flushleft}
\end{minipage}
}
  \caption{Mining Phase}
  \vspace{-0.2in}
   \label{fig:iteralg}
\end{figure}

\subsection{Implementation Detail}

In this section, we explain some of the implementation details of \alg. We use
Hadoop~1.1.2\footnote{http://hadoop.apache.org/releases.html} a open source
implementation of MapReduce framework written in Java. We use Java to write the
baseline mining algorithm as well as the map and the reduce function in the
preparation and the mining phase. We override the default input reader and
write a custom input reader for the preparation phase. To improve the execution
time of MapReduce job, we compress the data while writing them in HDFS. We
used global counter provided by Hadoop to track the stopping point of the
iterative mining task.

\subsubsection{MapReduce Job Configuration}

In this section, we provide a detail snapshot of the Map-Reduce job
configuration for \alg\ in Hadoop. Success of a job depends on the accurate
configuration of that job. Since the type of the value that is read by a mapper and emit
by a reducer is BytesWritable, \alg\ sets input and output format of each job as
\texttt{SequenceFileInputFormat} and \texttt{SequenceFileOutputFormat}.
Another job property, named
\texttt{mapred.task.timeout} also need to be set properly
for better execution of \alg. This parameter controls for how long the master
node waits for a data node to reply.  If the mining task that \alg\ commence is
computationally demanding, the default timeout which is 10 minutes may not be
enough. To be on the safe side, \alg\ sets the timeout of a job to 5 hour (300
minutes). \alg\ also sets the \texttt{mapred.output.compress} property to true. This
configuration lets the output of a MapReduce job to be compressed which
eventually decreases network load and improves the overall execution time. The
codec that is used for compression is \texttt{BZip2Codec}. \alg\ also increases the
heap size of a job using \texttt{mapred.child.java.opts} property. The same
configuration is used for both the preparation and mining phase of \alg.

\section{Experiments and Results}\label{sec:EXP}

In this section, we present experimental results that demonstrate the
performance of \alg\ for solving frequent subgraph mining task on large graph
datasets. As input, we use five real-world graph datasets which are taken from
an online source~\footnote{http://www.cs.ucsb.edu/∼xyan/dataset.htm} that
contains graphs extracted from the PubChem website~\footnote{
http://pubchem.ncbi.nlm.nih.gov}.  PubChem provides information on
biological activities of small molecules and the graph datasets from PubChem represent 
atomic structure of different molecules. In
Table~\ref{tab:real_dataset_stat}, we provide the name and statistics of these datasets.
We also create four synthetic datasets using a tool called
Graphgen~\cite{Cheng.Ke.ea:07}. The number of graphs in these datasets range
from 100K to 1000K and each graph contains on average $25-30$ edges.  We
conduct all experiments in a 10-node Hadoop cluster, where one of the node is
set to be a master node and the remaining 9 nodes are set to serve as data
node. Each machine possesses a 3.1 GHz quad core Intel processor with 16GB
memory and 1 TB of storage.

\begin{table}
\center
\begin{tabular}{|l|l|l|l|}\hline
Transaction        		& \# of Transactions 	& Average Size of  	\\
Graph dataset			&			& each transactions	\\\hline
NCI-H23   	 		& 40,353  		& 28.6  		\\\hline
OVCAR-8     			& 40,516  		& 28.1  		\\\hline
SN12C				& 40,532 		& 27.7   		\\\hline
P388  				& 41,472 		& 23.3			\\\hline
Yeast				& 79,601		& 22.8			\\\hline
\end{tabular}
\vspace{-0.in}
\caption{Real life biological Datasets}
\label{tab:real_dataset_stat}
\end{table}  

\subsection{Runtime of \alg\ for different minimum support}

\begin{figure}[!ht] 
  \centering
  \subfloat[]{    
    \includegraphics[width = 1.6in]{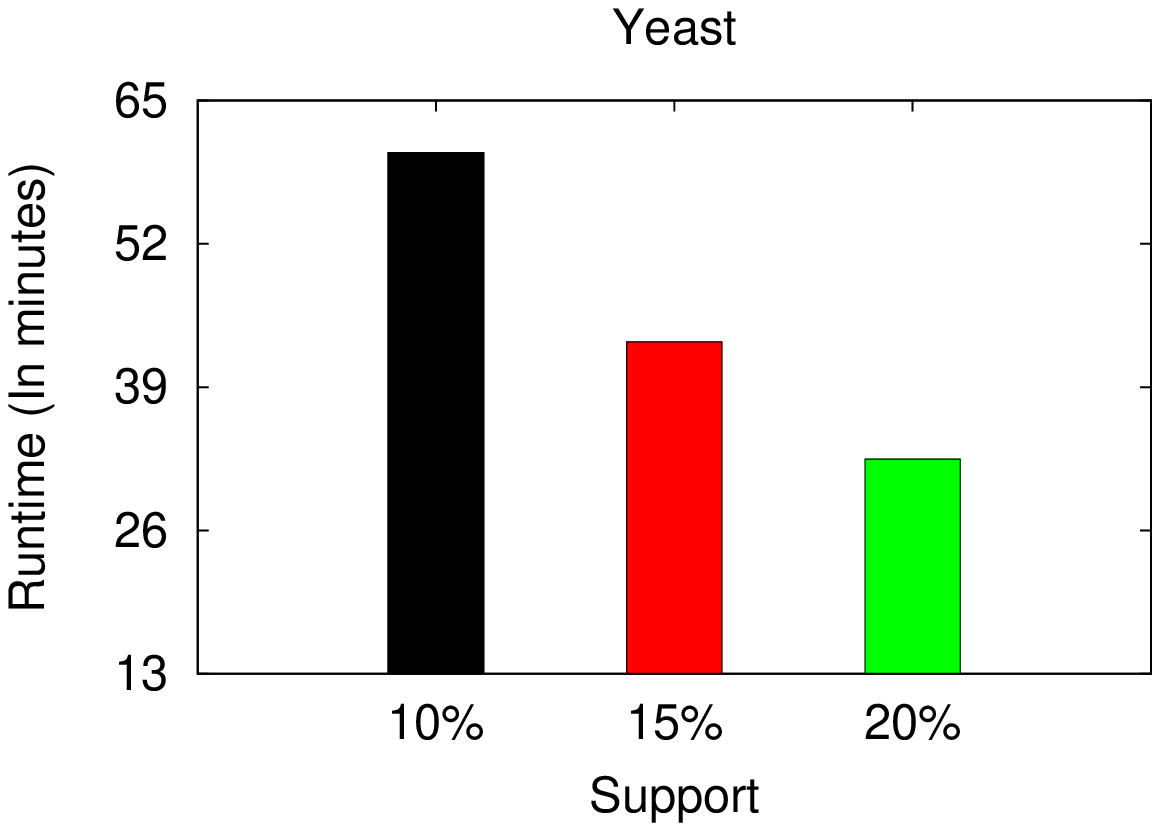}
  } 
  \subfloat[]{    
    \includegraphics[width=1.6in]{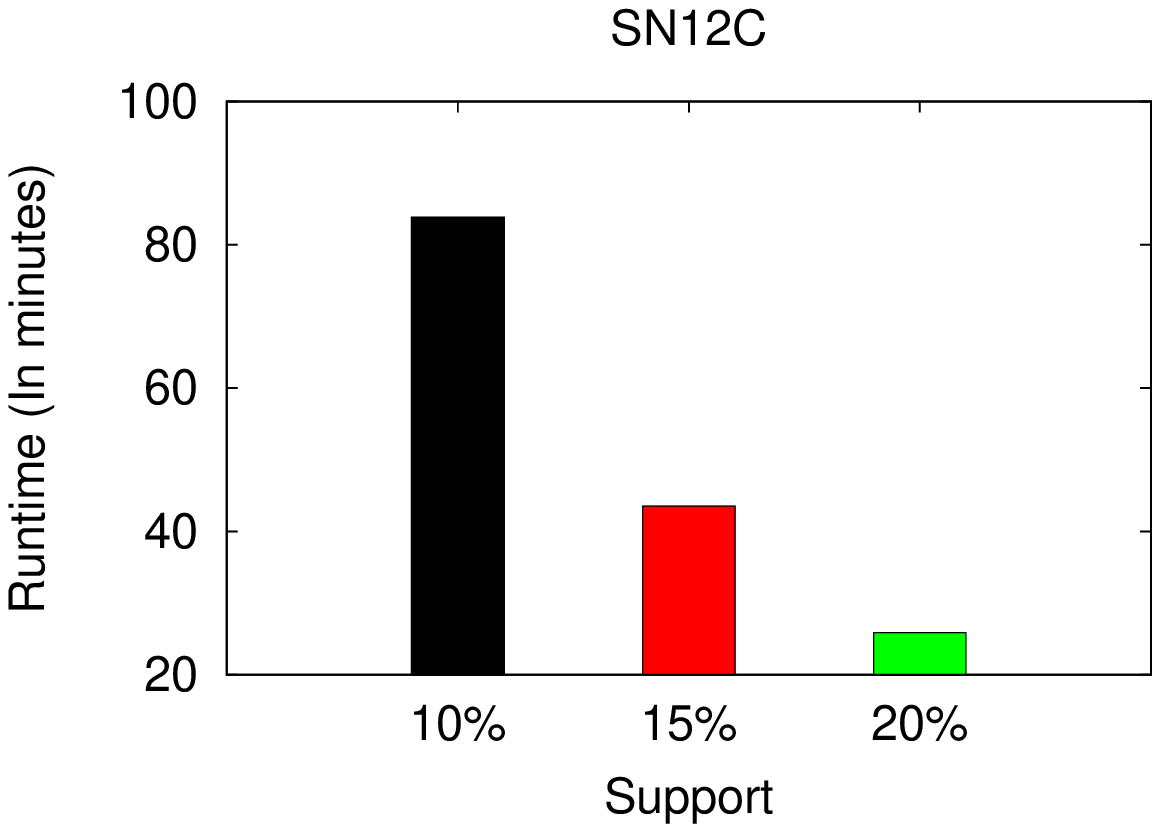}
  }\\ 
  \vspace{-0.2in}
  \subfloat[]{    
    \includegraphics[width=1.6in]{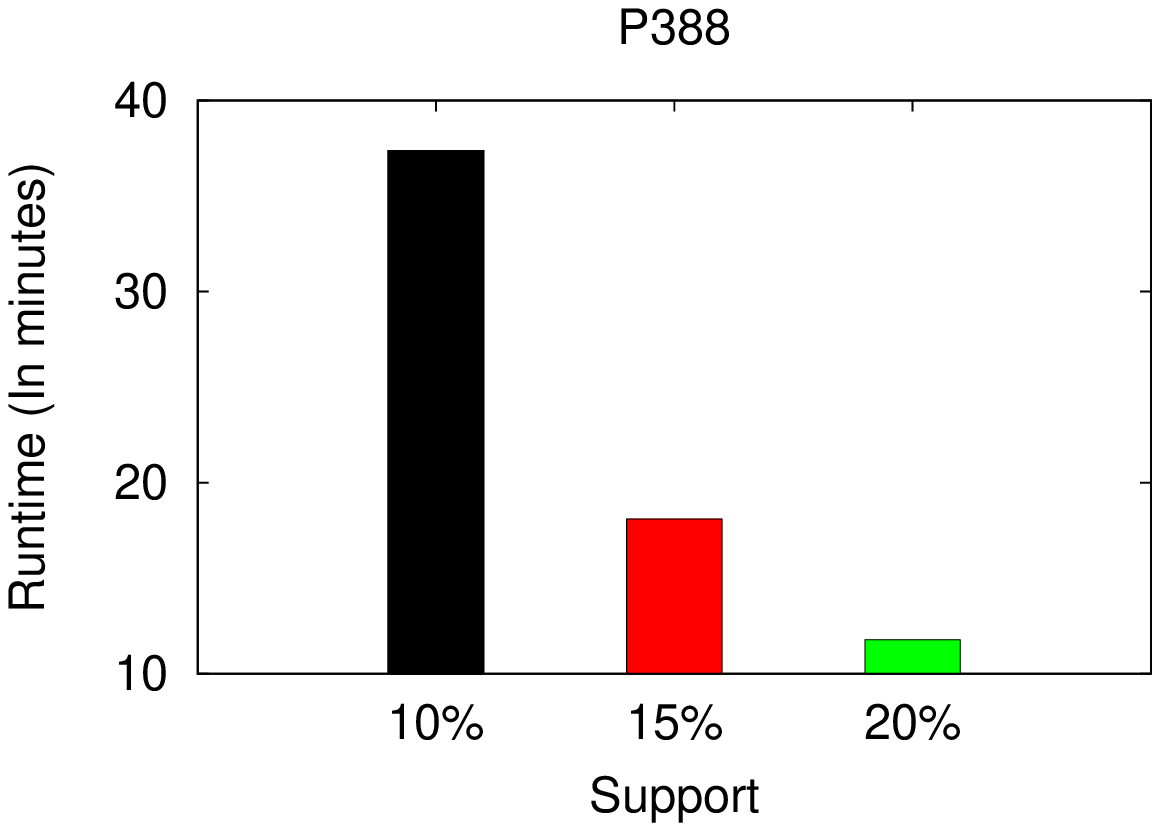}
  }
  \subfloat[]{    
    \includegraphics[width=1.6in]{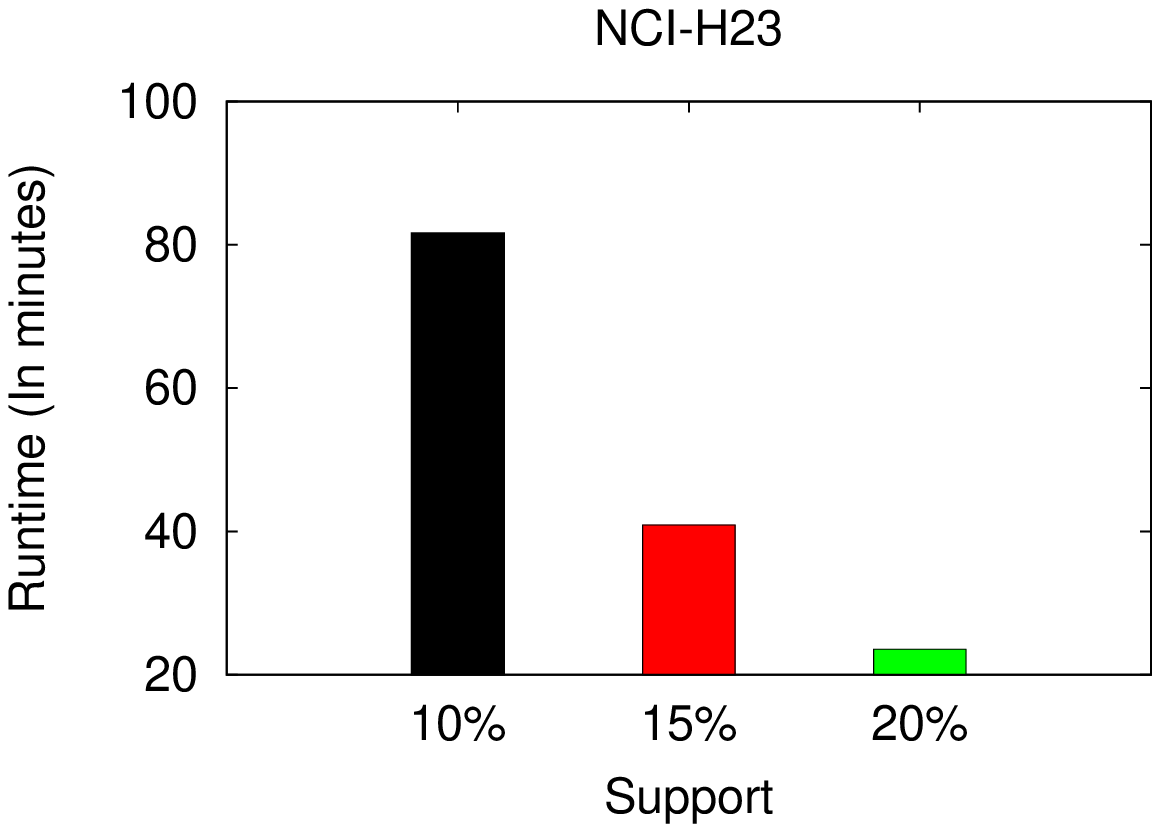}
  } \\ 
  \vspace{-0.2in}
  \subfloat[]{    
    \includegraphics[width=1.6in]{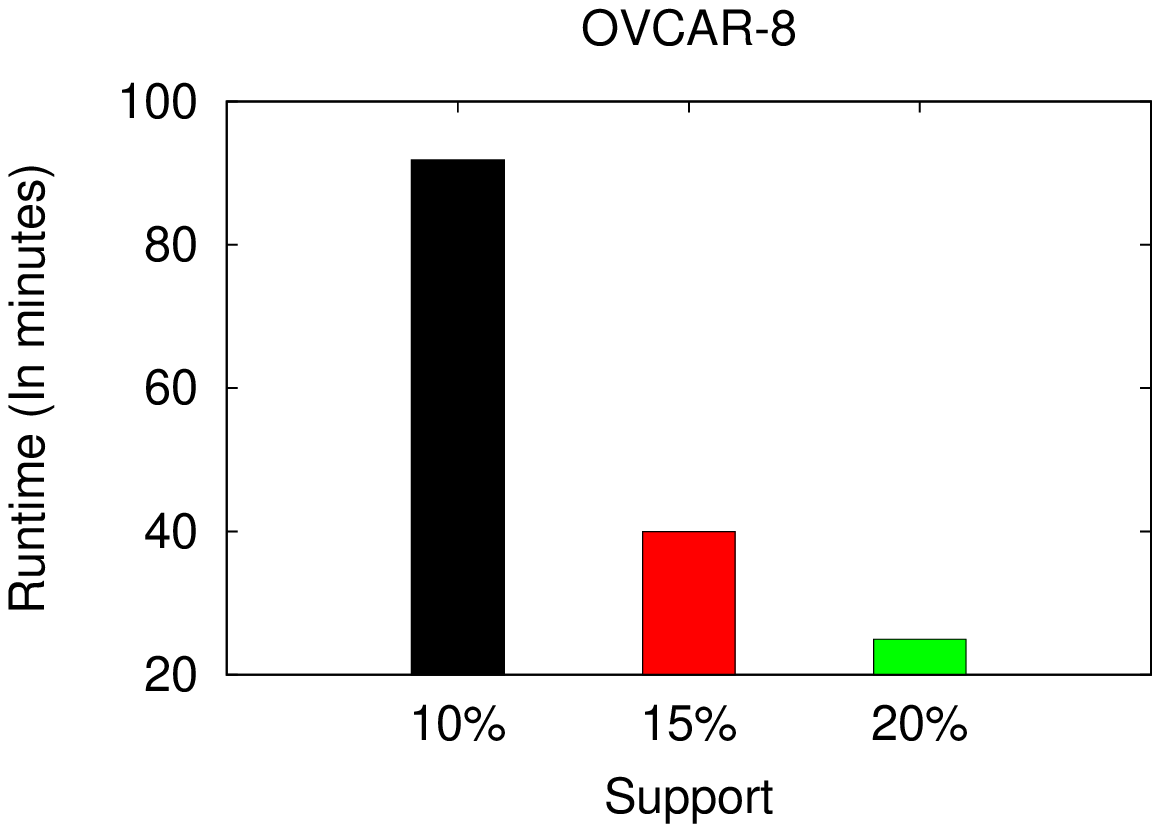}
  }
  \vspace{-0.1in}
  \caption{Bar plot showing the relationship between the minimum support threshold ($\%$) and the running time (in minutes) of \alg\ in (a) Yeast
 (b) NCI-H23 (c) P388 (d) SN12C and (e) OVCAR-8.}
 \vspace{-0.2in}
 \label{fig:support_vs_runtime}
\end{figure}

In this experiment, we analyze the runtime of \alg\ for varying minimum
support threshold. We conduct this experiment for biological datasets mention above. Here
we fix the number of data nodes to $9$ and keep track of the running time 
of \alg\ for minimum support thresholds that vary between $10\%$ to $20\%$. In
Figure~\ref{fig:support_vs_runtime}(a-e) we show the result. As expected, the
runtime decreases with the increase of minimum support threshold. 

\begin{table}
\center
\begin{tabular}{|l|l|l|l|}\hline
Number of transaction			& (runtime in minute)	\\\hline
100K 					& 27.4 			\\\hline
250K					& 69.1 			\\\hline
750K   	 				& 86.1  		\\\hline
1000K     				& 98.7  		\\\hline
\end{tabular}
\caption{Runtime of \alg\ on synthetic transaction graph datasets}
\vspace{-0.2in}
\label{tab:synthetic}
\end{table}

\subsection{Runtime of \alg\ for different number of database graphs} For this
experiment, we generate 4 synthetic datasets each having 100K to 1000K input
graphs. Each of these graphs has 25 vertices and their edge density is around
$0.5$ or smaller. Table~\ref{tab:synthetic} shows the runtime of \alg\ over
these datasets for $30\%$ minimum support threshold using $1000$ partitions for
each datasets..

\subsection{Runtime of \alg\ on varying number of data nodes}

\begin{figure}[!ht] 
\vspace{-0.2in}
  \centering
  \subfloat[]{    
    \includegraphics[width = 1.6in]{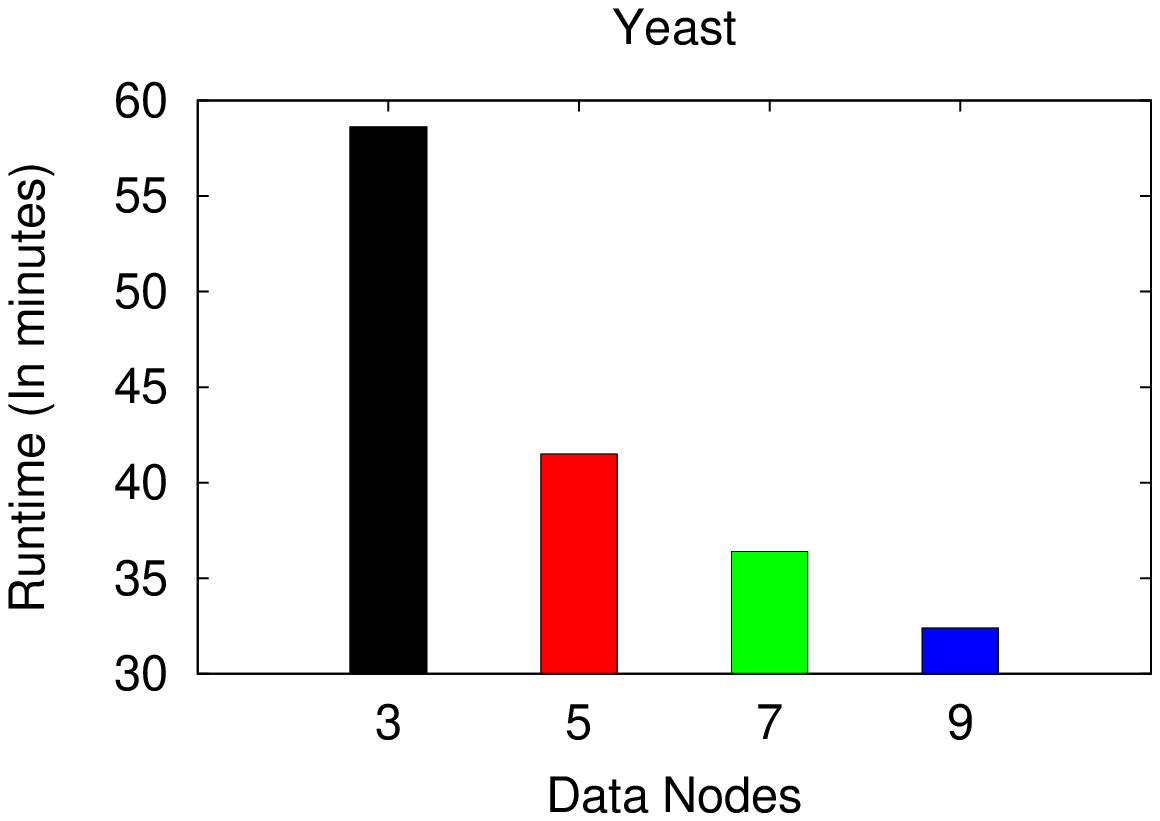}
  } 
  \subfloat[]{    
    \includegraphics[width=1.6in]{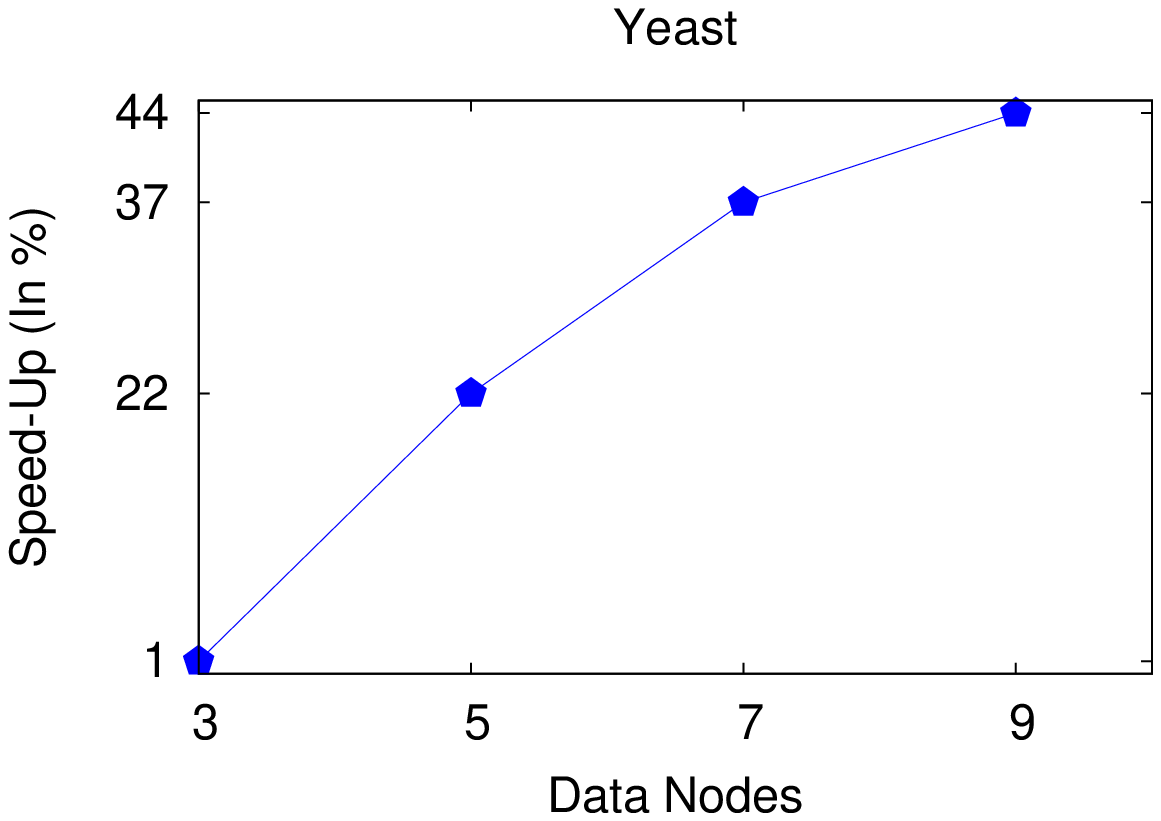}
  }
  \vspace{-0.1in}
  \caption{Relationship between the execution time and the number of data nodes: (a) Bar plot shows the
execution time; (b) Line plot shows the speedup with respect to the execution time using 3 data nodes configuration}
  \vspace{-0.2in}
 \label{fig:nodes_vs_runtime}
\end{figure}

In this experiment, we demonstrate how \alg's runtime varies with the number of
active data nodes (slaves). We use the Yeast dataset using 20\% minimum
support threshold. We vary the count of data nodes among 3, 5, 7 and 9 and
record the execution time for each of the configurations.  As shown in
Figure~\ref{fig:nodes_vs_runtime}(a) the runtime reduces significantly with
an increasing number of data nodes. In Figure~\ref{fig:nodes_vs_runtime}(b) we
plot the speedup that \alg\ achieves with an increasing number of data nodes, with
respect to the 3-data-nodes configuration. We can see that the speedup 
increases almost linearly except for the last data point. 

\subsection{Runtime of \alg\ for varying number of Reducer}

The number of reducers plays an important role in the execution of MapReduce
job in Hadoop.  While writing data (output) in HDFS, a MapReduce job follows a
convention of naming the output file with the key word ``part''.
Reducer count determines how many ``part'' files will be generated to hold
the output of a job. If the number of reducer is set to 1, entire output will be
written in a single file. Since \alg\ is an iterative algorithm, where output
of the current job is used as an input of the next job, the number of reducer
has a significant effect on the execution time of \alg\ .  If we set reducer count to a
small value then there will be fewer number of output files that are large in
size; these large files will be a burden over the network when they are
transfered between data nodes. On the other hand, large number of reducers
might create many output files of zero size (reducer is unable to output any
frequent pattern for the next stage Mapper). These zero size output files will
also become an overhead for the next stage mappers as these files will
still be used as input to a mapper. Note that, loading an input file is costly
in Hadoop. 

\begin{figure}[!ht] 
\vspace{-0.2in}
  \centering
  \subfloat[]{    
    \includegraphics[width = 1.6in]{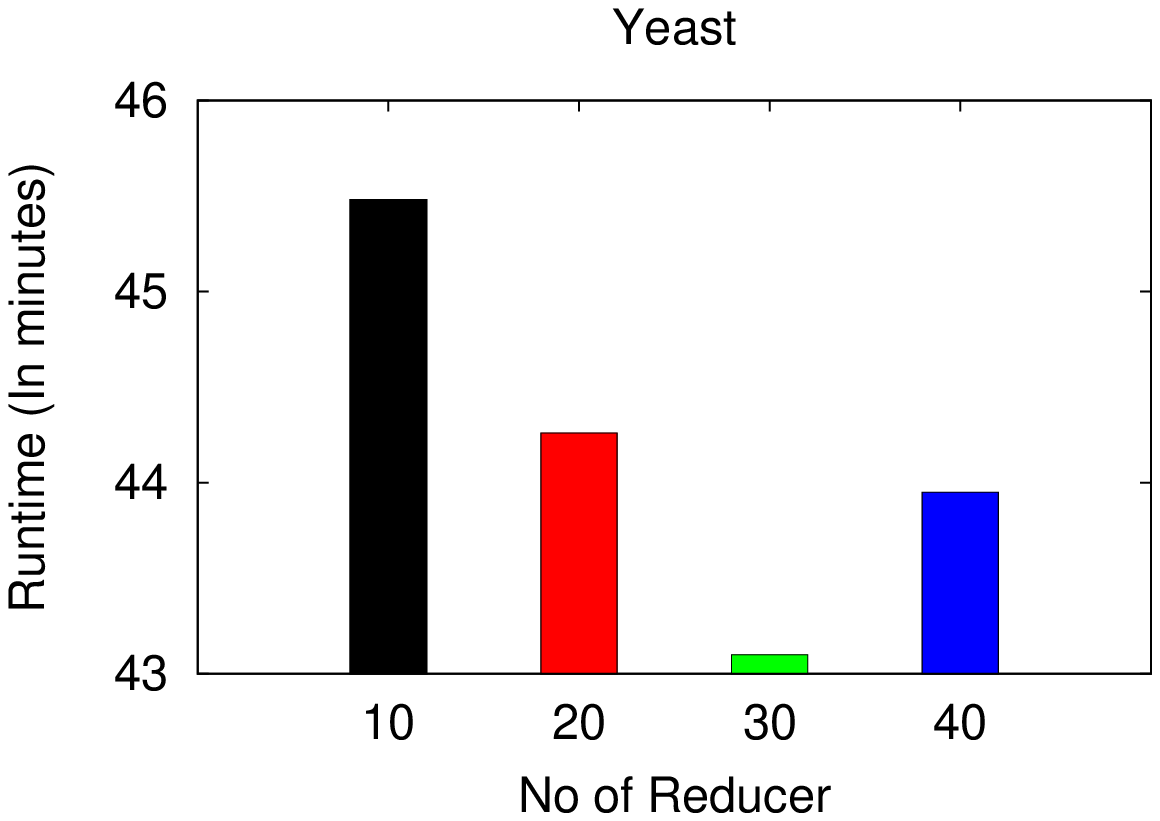}
  } 
  \subfloat[]{    
    \includegraphics[width=1.6in]{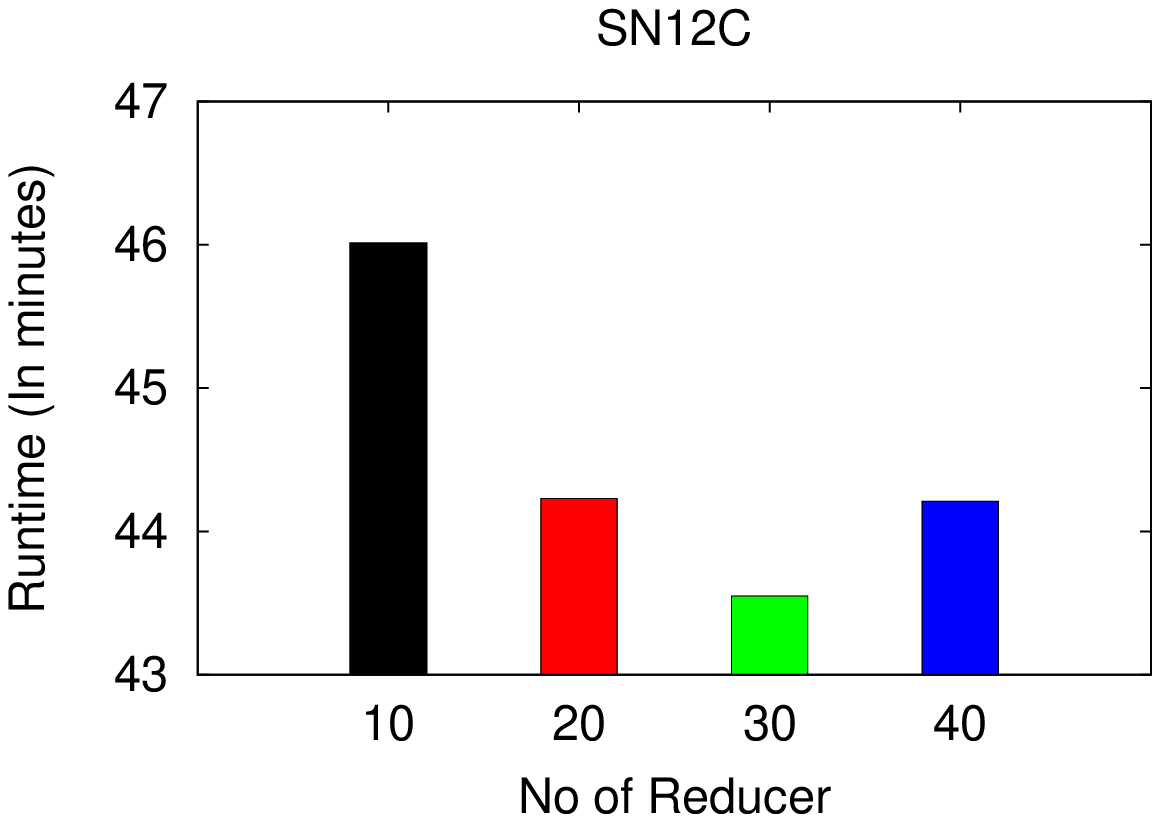}
  }\\ 
  \vspace{-0.2in}
  \subfloat[]{    
    \includegraphics[width=1.6in]{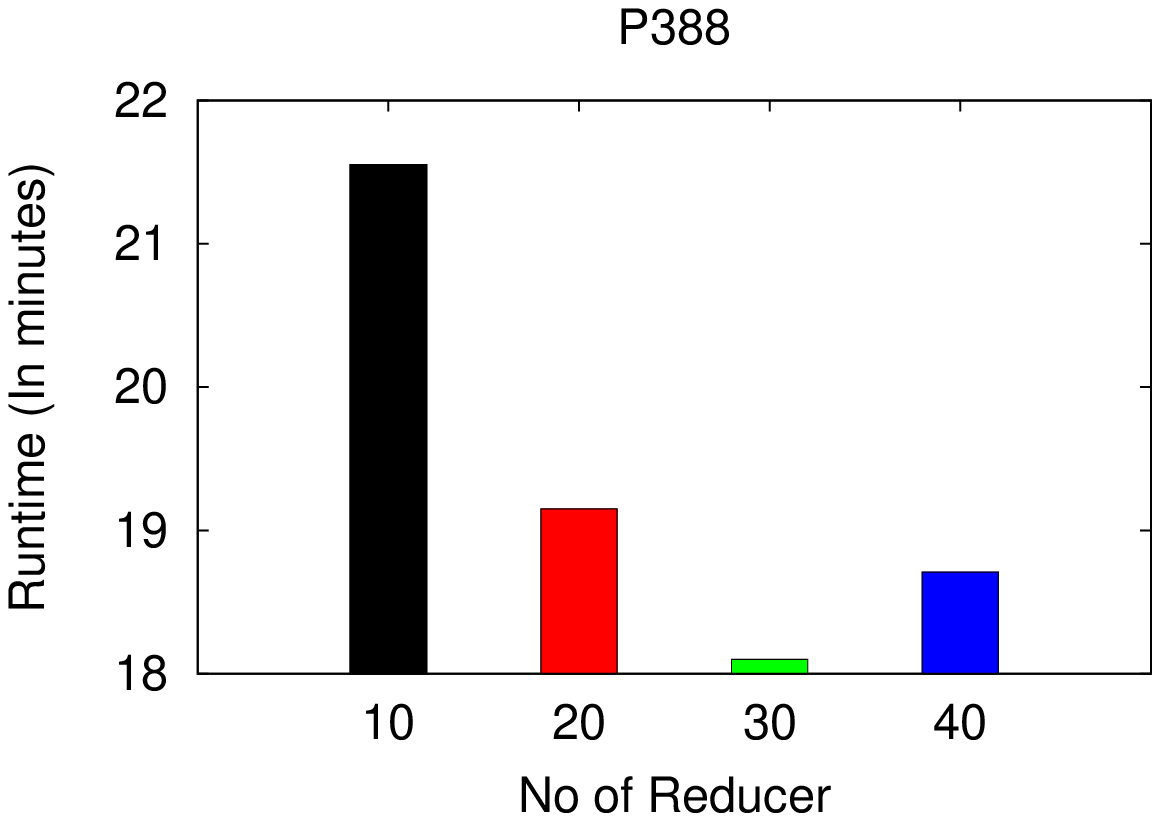}
  }
  \subfloat[]{    
    \includegraphics[width=1.6in]{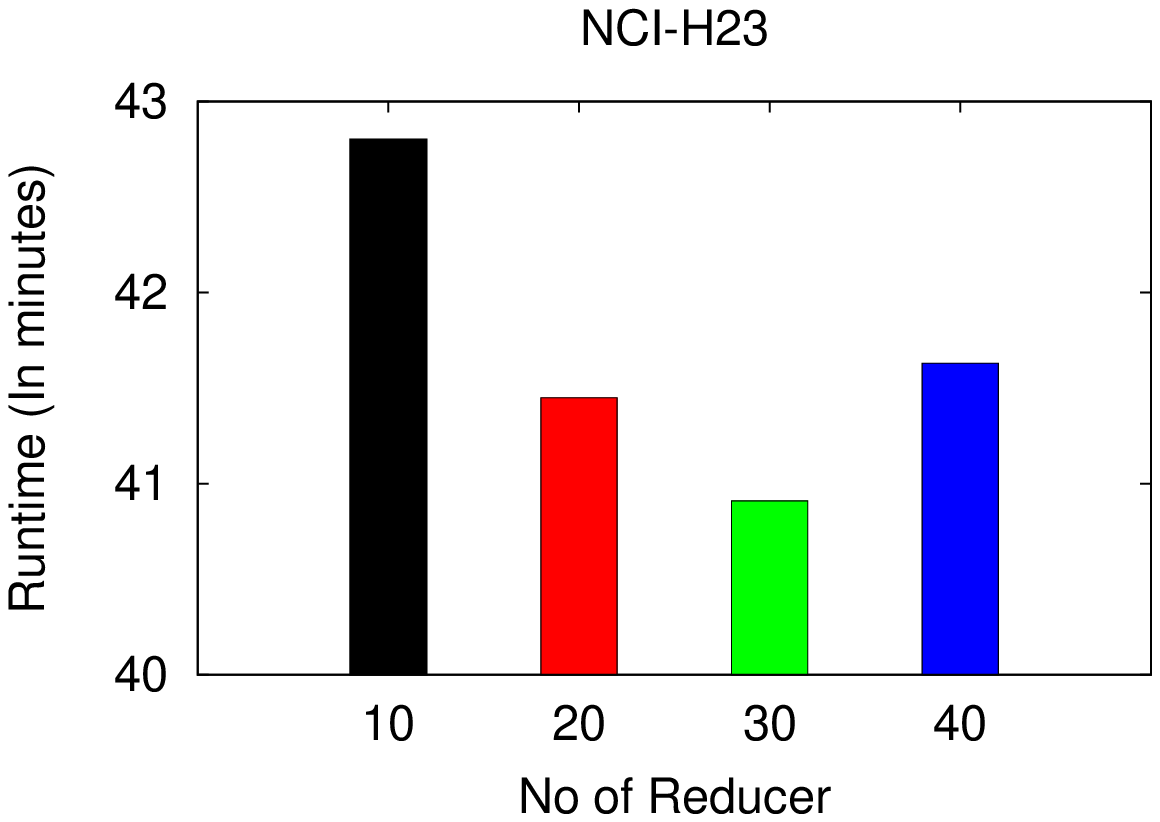}
  } \\ 
   \vspace{-0.2in}
  \subfloat[]{    
    \includegraphics[width=1.6in]{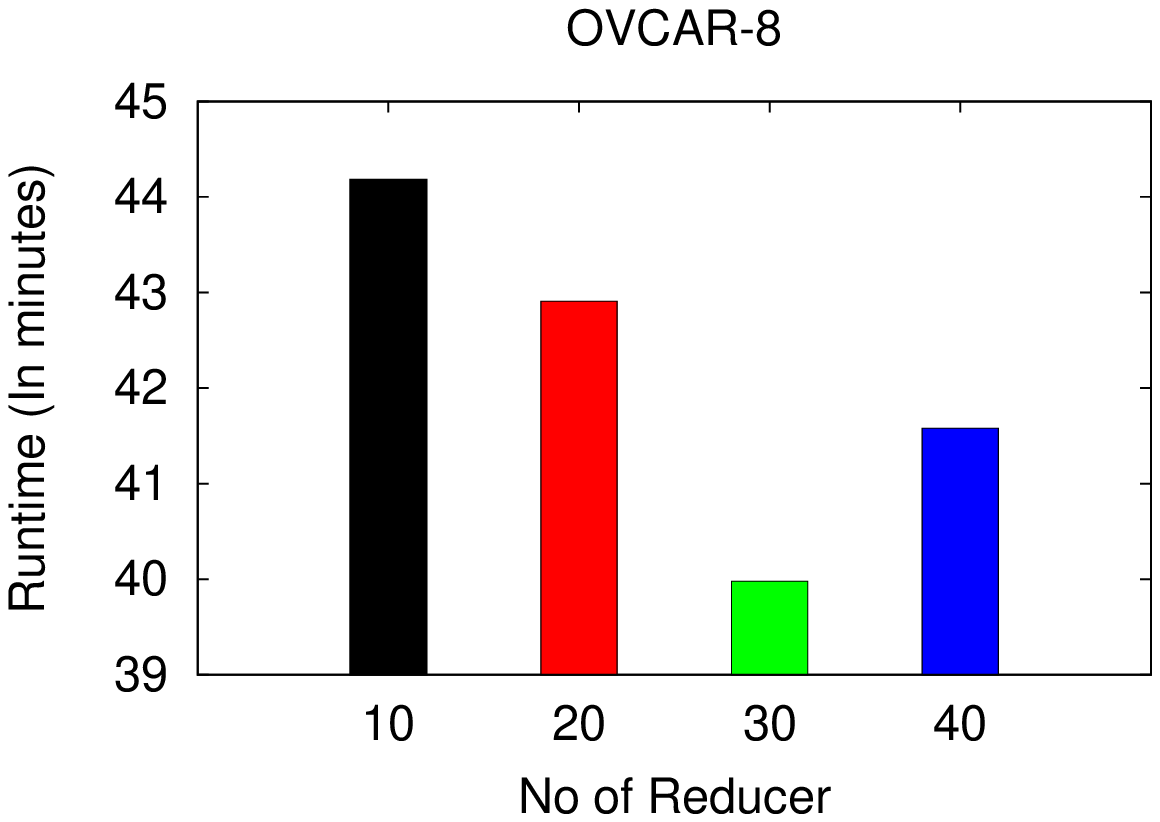}
  }
  \vspace{-0.14in}
  \caption{Bar plot showing the relationship between the number of Reducer and the running time of \alg\ in (a) Yeast,
 (b) NCI-H23, (c) P388, (d) SN12C, and (e) OVCAR-8.}
 \vspace{-0.2in}
 \label{fig:Reducer_vs_runtime}
\end{figure}

In this experiment, We measure the runtime of \alg\ for various configurations
such as, 10, 20, 30 and 40 reducers. We run the experiment on all 5 real-life
datasets for 20\% minimum support threshold.
Figure~\ref{fig:Reducer_vs_runtime}(a-e) shows the relationship between
execution time and the number of reducers using bar charts. As we can see from all
the charts, 30 is the best choice for the number of reducers in our cluster
setup. This finding is actually intuitive, because we have 9 data nodes each
having 4 reduce slots (4 core processor), yielding 36 processing units.  So
keeping a few units for system use, 30 is the best choice for the number of
reducers.

\begin{figure}[!ht] 
\vspace{-0.2in}
  \centering
  \subfloat[]{    
    \includegraphics[width = 1.6in]{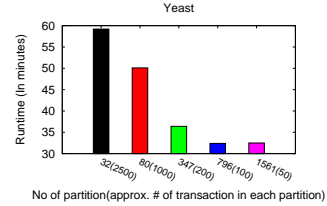}
  } 
  \vspace{-0.14in}
  \caption{Bar plot showing the relationship between the partition count and the running time of \alg\ for Yeast dataset.}
  \vspace{-0.2in}
 \label{fig:partition_vs_time}
\end{figure}

\subsection{Runtime of \alg\ for different partition sizes}

An important requirement for achieving the optimal performance of a Hadoop
cluster is to find the appropriate number of mappers. In such a cluster, the
setup and scheduling of a task wastes a few seconds. If each of the
tasks is small and the number of tasks is large, significant amount of time
is wasted for task scheduling and setup. So the experts advise that
a mapper should execute for at least 40
seconds~\footnote{http://blog.cloudera.com/blog/2009/12/7-tips-for-improving-mapreduce-performance/}.
Another rule of thumbs is that the number of partition should not over-run the
number of mappers.

In our cluster setup, we have 9 data nodes, each with 4 cores that allow us to
have 36 mappers.  Then, the perfect number of input partitions should be less
than or equal to $36$. But this rules does not fit well for frequent subgraph
mining task.  For example, the Yeast dataset has close to $80,000$ graphs and
if we make $30$ partitions, then each partition ends up consisting of $2666$
graphs. Since, frequent subgraph mining is an exponential algorithm over its
input size, performing a mining task over these many graphs generally becomes
computationally heavy.  In this case, the map function ends up taking time much
more than expected. As a result the optimal number of partitions for an FSM
task should set at a much higher value (compared to a tradition data analysis
task) so that the runtime complexity of the mappers reduces significantly. Note
that, the higher number of partitions also increases the number of key-value
pairs for a given patterns which should be processed by the reducers. However,
the performance gain from running FSM over small number of graphs supersedes
the performance loss due to the increased number of key-value pairs. This is
so, because the gain in execution time in the mappers follows an exponential
function, whereas the loss in execution time in the reducers and the data
transmission over the network follow a linear function.

The following experiment validates the argument that we have made in the above
paragraph. In this experiment, we run \alg\ on Yeast dataset for different
number of partitions and compare their execution time.
Figure~\ref{fig:partition_vs_time} shows the result using bar charts. The
charts show that as we increase the partition count, the performance keeps
improving significantly until it levels off at around 1000 partitions. When the
partition count is 1561, there is a slight loss in \alg's performance compared
to the scenario when the partition count is 796.  The strategy of finding the
optimal number of partitions depends on the characteristics of input graphs that
control the complexity of the mining task, such as the density, number of edges,
number of unique labels for vertices and edges in an input graph and so on.

\subsection{Comparison of \alg\ with an existing algorithm}

As discussed in section~\ref{sec:RW}, \cite{Hill.Srichandan.ea:2012} proposed a
MapReduce based graph mining method which is inefficient due to its poor design
choices.  In this experiment, we  compare the execution time of \alg\ with
that of Hill et al.'s~\cite{Hill.Srichandan.ea:2012} implementation that we
obtain from the authors. We run both the algorithms on three real world
biological dataset for $40\%$ minimum support. Table~\ref{tab:comparison}
compares the execution time between the two algorithms. As we can see
\alg\ performs significantly better that the other algorithm.

\begin{table}
\center
\begin{tabular}{|l|l|l|l|}\hline
Dataset					& \alg\ (min) 	& Hill et. al's\cite{Hill.Srichandan.ea:2012}(min)	\\\hline
Yeast 					& 16.6 		& 109.2							\\\hline
P388					& 8.3 		& 57.5							\\\hline
NCI-H23   	 			& 17.5  	& 117.2							\\\hline
\end{tabular}

\caption{Runtime (in minutes) comparison between \alg\ and Hill et. al's\cite{Hill.Srichandan.ea:2012} on three biological graph datasets}
\vspace{-0.1in}
\label{tab:comparison}

\end{table} 
\subsection{Effect of partition scheme on runtime}

In this experiment, we analyze how the partition heuristics discussed in
Section~\ref{sec:partition}  affects the execution time of \alg.  For all
datasets, We fixed minimum support to $20\%$, number of data nodes to $9$ and
the number of reducer to 30. The result is shown in
Table~\ref{tab:partition_scheme}.  We find that for schema 2, which balances
the total number of edges in each partition, performs somewhat better than
schema 1, which only balances the number of graphs.  Note that, in the
real-life graph datasets most of the graphs have similar number of edges, so
the partitions using schema 1 and schema 2 have similar balance factors. But,
for datasets where the input graphs have highly different number of vertices
and edges, the schema 2 is obviously a better choice. To investigate further,
we build a synthetic dataset of $50K$ graphs where average length of half of the
graphs is around $15$ and for the other half it is around $30$. We found that for this unbalanced
dataset, scheme 1 takes about 30\% more time than the schema 2 (see the last
row of the Table~\ref{tab:partition_scheme}).

\begin{table}
\center
\begin{tabular}{|l|l|l|l|}\hline
Transaction        		& Scheme 1 		& Scheme 2  	\\
Graph dataset			& (runtime in min)	& (runtime in min)	\\\hline
Yeast 				& 32.5 			& 30.3  		\\\hline
SN12C				& 25.8 			& 23.7   		\\\hline
P388   	 			& 11.7  		& 10.2  		\\\hline
NCI-H23     			& 23.5  		& 22  			\\\hline
OVCAR-8  			& 24.9 			& 23.4			\\\hline
Synthetic                       & 22.9                  & 17.1                 	\\\hline

\end{tabular}
\caption{Runtime of \alg\ on different partition schemes}
\vspace{-0.2in}
\label{tab:partition_scheme}

\end{table}  

\section{Conclusions}\label{sec:CON}

In this paper we present a novel iterative MapReduce based frequent subgraph
mining algorithm, called \alg. We show the performance of \alg\ over real
life and large synthetic datasets for various system and input configurations. 
We also compare the execution time of \alg\ with an existing method, which
shows that \alg\ is significantly better than the existing method.

%

\bibliographystyle{IEEEtran}
\balance
\bibliography{mybib}  

\end{document}